\documentclass[a4paper,fleqn,usenatbib]{mnras}

\usepackage{newtxtext,newtxmath}

\usepackage[T1]{fontenc}
\usepackage{ae,aecompl}

\usepackage{graphicx}	
\usepackage{amsmath}	

\title[Recurrent Nova U Scorpii]{The Recurrent Nova U Scorpii from the 2010.1 to 2022.4 Eruptions; the Missed Eruption of 2016.78$\pm$0.10 and the Critical  Complex Period Changes}

\author[B. E. Schaefer]{
Bradley E. Schaefer$^{1}$\thanks{E-mail: schaefer@lsu.edu},
\\
$^{1}$Department of Physics and Astronomy, Louisiana State University, Baton Rouge, Louisiana, 70820, USA\\
}

\pubyear{2021}

\begin{document}
\label{firstpage}
\pagerange{\pageref{firstpage}--\pageref{lastpage}}
\maketitle

\begin{abstract}

U Sco is a recurrent nova with eleven observed eruptions, most recently in 2010.1 and 2022.4.  I report on my program (running since 1989) of measuring eclipse times and brightnesses of U Sco in quiescence, from 2010 to 2022.  The orbital period suddenly increased by $+$22.4$\pm$1.0 parts-per-million across the 2010.1 eruption.  This period change is greater than the near-zero period change ($+$3.9$\pm$6.1 parts-per-million) across the 1999.2 eruption.  This period change cannot come from any of the usual mechanisms, whereas the one remaining possibility is that the period changes are dominated by the little-known mechanism of the nova ejecting asymmetric shells.  From 2010.1 to 2016.78, the $O-C$ curve showed a steady period change that was large, with $\dot{P}$=($-$21.0$\pm$3.2)~$\times$~10$^{-9}$.  This is greatly higher than the steady period changes in the two previous inter-eruption intervals ($-$3.2$\pm$1.9 and $-$1.1$\pm$1.1 times 10$^{-9}$).  This large, variable, and negative $\dot{P}$ apparently comes from magnetic braking of the companion star's rotation.  Starting in 2016.9$\pm$0.6, the $O-C$ curve showed a strong kink that is a unique characteristic of the sudden period change ($+$35.4$\pm$7.1 parts-per-million) across a nova event.  The brightness in quiescence after 2010.4 shows that the white dwarf accreted the trigger mass for the next nova event in the year 2017.1$\pm$0.6.  Photometric records show the only possible time for the eruption to peak (such that its total duration of 60 days was undetectable by any observation) is during a 75 day interval inside the 2016 solar gap, thus constraining the missed eruption to 2016.78$\pm$0.10.
 
\end{abstract}




\section{INTRODUCTION}

U Scorpii (U Sco) is a famous recurrent novae (RN), out of a total of 10 known RNe in our entire Milky Way galaxy (Schaefer 2010).  U Sco is a cataclysmic variable (CV) where thermonuclear nova eruptions occur with a recurrence time-scale of $\sim$10 years.  U Sco has eleven observed eruptions in 1863, 1906, 1917, 1936, 1945, 1969, 1979, 1987, 1999, 2010 and now in 2022.  The time between eruptions, $\delta Y$, was as low as 7.9 years.  

U Sco is a very-fast high-energy nova with extreme properties.  (See Schaefer 2010 for a comprehensive review.)  With no precursor in the pre-eruption light curve, the nova goes from its usual quiescent level of $V$=17.6, rises in just a few hours to peak brightness of $V$=7.5, has a very short peak of only a day or so, and fades by 3 magnitudes from peak in the record-breaking 2.6 days.  The entire eruption from quiescence-to-peak-to-quiescence lasts just 60 days.  The nova spectrum is of the He/N class, displays the high-energy He {\rm II} emission lines, and has expansion velocities of 5700 km/s as measured from the FWHM of the H$\alpha$ line.  Spectra in quiescence reveal absorption lines from a G0 companion star.  In 1989, I discovered deep eclipses with a period of 1.23 days.

The deep eclipses of U Sco allow for a very accurate measure of the orbital period $P$, and for the changes in the period.  The period changes carry a variety of unique information that directly tests the most important issues of CV evolution, hence U Sco eclipse times have a high importance.  With this realization, I started a concerted and persistent program of measuring U Sco eclipse times, fully in the realization that this would be a lifetime task.  My primary work for U Sco study has been to measure eclipse times and the changes in the orbital period (Schaefer 1990; 2010; Schaefer \& Ringwald 1995, Schaefer et al. 2010; 2011).  This has expanded to include all galactic RNe (Schaefer 1990; 2009; 2010; 2011; Schaefer et al. 1992; 2013) and normal galactic novae (Schaefer 2020a; 2020b; Schaefer \& Patterson 1983; Schaefer et al. 2019).  Applications have been testing whether RNe and novae can become Type Ia supernovae, testing the Hibernation scenario of CV evolution, and testing the Magnetic Braking model of CV evolution.  The photometry reported in this paper is a continuation of my long series of eclipse times of U Sco, other RNe, and other classical novae.  

Schaefer (2005) invented a method to predict the dates of upcoming eruptions of RNe, based only on the brightness history in quiescence, and predicted the next eruption of U Sco for the year 2009.3$\pm$1.0.  Based on this prediction and the importance of U Sco, starting in 2008, A. Pagnotta (now at the College of Charleston) and myself organized a program of intense observations.  The first part of this program was intensive monitoring of U Sco by many observers for the eruption, with an average of a dozen times spread widely through each day, all with a rapid alert mechanism with the 24-hour-every-day clearinghouse provided by the American Association of Variable Star Observers (AAVSO).  Further advance work was the writing and acceptance of many Target-of-Opportunity proposals for spacecraft and observatories.  The eruption was discovered on 2010 January 28, independently by B. J. Harris (AAVSO observer code HBB) and S. Dvorak (observer code DKS), both AAVSO observers in Florida.  The alert went out to the world within minutes, followed by a massive observing campaign worldwide.  The 2010 U Sco eruption was observed in radio, far-infrared, near-infrared, optical, ultraviolet, X-ray, and $\gamma$-ray bands.  Six different spacecraft observatories were used.  A dozen eclipse times were measured from the last 47 days of the eruption tail.  For photometry, a total of 36,776 magnitudes were measured evenly across the 60 day of the eruption, with an average of one magnitude every 140 seconds for the entire 60 days straight.  This was the first time that any nova had ever been looked at with time series photometry of any duration, and two entirely unexpected new phenomena were discovered.  (One is the still-inexplicable hour-long large-amplitude flares that appeared only around the transition time roughly one week after peak, and the other is the deep aperiodic eclipses seen in the later parts of the eruption tail, presumably caused by raised edges splashed on to the outer edge of the reforming accretion disc.)  A summary and review of this campaign, along with the detailed science conclusions is presented in Pagnotta et al. (2015).  More specialized studies include the pre-eruption light curve and discovery (Schaefer et al. 2010), eclipse times (Schaefer 2011; Schaefer et al. 2011), optical spectroscopy (Diaz et al. 2010; Mason 2011; Kafka \& Williams 2011; Mason et al. 2012), X-ray measures (Ness et al. 2012; Orio et al. 2013; Takei et al. 2013), ultraviolet measures (Page et al. 2013), and infrared measures (Banerjee et al. 2010).  With this intensive campaign, U Sco in 2010 became the all-time best-observed nova event.

During the later parts of the year 2008 and then 2009, with all the advance preparation for the intensive upcoming campaign, Pagnotta and myself became horrified with the possibility of a `nightmare scenario'.  This worst case was if U Sco would have its entire eruption lost within a `solar gap' with no observations, such that we would not even know that the eruption had been missed.  The solar gap is the time interval each year when the Sun passes too close to the nova to allow observations, with the solar gap centred on 29 November each year with the solar conjunction of 3.6$\degr$.  U Sco is largely unobservable for an 80--150 day interval, depending how close observers push to watch the nova low in the western sky soon after dusk and push to catch the nova low in the eastern sky before dawn.  U Sco in eruption is significantly above its quiescent level only for 60 days (Pagnotta et al. 2015), and it is easily possible for an eruption that peaks soon after the season's last observation would be completely missed.  An eruption peaking within 60 days (or less) of the first observation of the next season would be discovered as a fading tail, even though the peak itself was missed.  (Exactly this happened in 1969, when the all-time-greatest variable star observer A. Jones and the legendary F. Bateson caught the tail at the end of an eruption in the dawn skies of New Zealand.)  Each year, there is a 20--90 day window over which an eruption could peak and yet remain completely invisible because the light curve before and after the solar gap shows U Sco at its normal quiescent level.  (And this is what happened for the missed eruptions around the years 1927 and 1957.)  Every individual eruption has a roughly 6--25 per cent chance of being entirely missed. 

This `nightmare scenario' is difficult to handle.  Had this worst case occurred in late-2008 or late-2009, the collaboration organized by Pagnotta and myself would have been kept on high alert, perhaps for years, writing target-of-opportunity proposals, and intensively monitoring U Sco for the next eruption that would not come for many years.  In this nightmare scenario, there is nothing that anyone could have done, and no one to blame.  Part of the solution is to recognize this case, even if only years later.  Yet, how can anyone realize that U Sco went through an eruption if the entire event was not observed inside a solar gap?  U Sco does not have an extended photometric tail in its eruption light curve to serve as a tell-tale.  U Sco does not have any spectral signature of a recent eruption.  U Sco does not have any expanding nova shell that can be spotted in subsequent months.  U Sco does not have detectable radio emission with its long duration that could be spotted long after the optical eruption is ended.  Our best solution, requiring a long wait of a few years, was that a missed eruption can be recognized by a kink in the $O-C$ curve.  Both observation and theory showed that nova will suffer a sudden sharp period change (by $\Delta P_{nova}$) across an eruption (Schaefer 2020b).  In the years following the missed eruption, a steady stream of eclipse timings would reveal a kink in the $O-C$ as the hallmark of a missed eruption.  That is the best that anyone can do.  Until the required $O-C$ kink is discovered, there is no proof of a missed eruption.

This nightmare scenario came about after the 2010.1 eruption.  With the eruption in 2010.1, the next eruption is expected around 2020, or perhaps as early as 2010.1+7.9=2018.0, or possibly earlier.  In 2018, as preparation for a second massive observing campaign, I collected U Sco magnitudes in quiescence, and realized that the nova had been anomalously bright after 2010.4.  This pointed to a short $\delta Y$, with an eruption that might have already been missed.  But this analysis does not constitute proof.  To get the proof, I tried two methods.  The first method was to seek deeply and worldwide for some previously overlooked measure showing U Sco significantly above its quiescent level, from 2015 to 2018, trying to push into the solar gaps as much as possible.  This included getting images from spacecraft with wide-field images from coronagraphs staring at the Sun, with their possibility of spotting U Sco near maximum light in the weeks around solar conjunction.  The second method was to obtain many eclipse timings, spread out over the years, to find a kink in the $O-C$ curve, as belated proof of a missed eruption.

This paper tells two intertwined stories, the first being the lost eruption in the year 2016.78$\pm$0.10, and the second being the front-line science returned from the measured period changes of U Sco.  Further, after the original submission of this paper to $MNRAS$, U Sco started another eruption on 2022 June 7.  This makes my study of U Sco in quiescence from 2010 to 2022 as being bounded by two well-observed eruptions.

\section{QUIESCENT LIGHT CURVE AFTER 2010.4}

It is important to have explicit light curves permanently available in the literature, as there are a variety of way for future analysts to look at the data, to re-use the old data for new analyses, and to test out now-unrecognized procedures.  For the 2010 eruption, long listings of optical, X-ray, and infrared light curves are presented in full by Pagnotta et al. (2015) and Schaefer et al. (2011).  Complete light curves of all magnitudes for all eruptions from 1863 to 1999 are presented in Schaefer (2010).  Schaefer (2005; 2010) and Schaefer et al. (2010) present the whole history of brightness measures of U Sco in quiescence  before 2010, while listing all magnitudes.  In Sections 2 and 3, I will explicitly list all of the photometry from 2010.4--2022.3.

\begin{table*}
	\centering
	\caption{U Sco photometry data sources after 2010.4}
	\begin{tabular}{llllllr}
		\hline
		Source & Reference & UT Start Date  & Start JD   &   End JD   &   Band   &   Number\\
		\hline
AAVSO	&	[1]	&	2010 Jun 1	&	2455349	&	2459655	&	{\it B, V, visual, CV, CR}	&	1915	\\
CRTS	&	[2]; Drake et al. (2009; 2014)	&	2010 Jun 5	&	2455352	&	2456459	&	{\it CV}	&	152	\\
Pan-STARRS	&	[3]; Chambers et al. (2016)	&	2010 Aug 23	&	2455431	&	2456887	&	{\it grizy}	&	54	\\
DECam	&	[4]	&	2013 Mar 1	&	2456352	&	2457166	&	{\it r, i, VR}	&	24	\\
CTIO	&	This paper	&	2013 May 17	&	2456429	&	2457623	&	{\it I}	&	276	\\
K2 Cycle 2	&	[5]; GO2007 PI Schaefer	&	2014 Aug 25	&	2456894	&	2456972	&	Kepler	&	106779	\\
ATLAS	&	[6]; Tonry et al. (2018), Heinze et al. (2018)	&	2015 Jul 25	&	2457228	&	2457923	&	cyan, orange	&	134	\\
CSS	&	[7]	&	2016 Jun 2	&	2457541	&	2458584	&	{\it CV}	&	77	\\
LASCO	&	[8]	&	2016 Nov 20	&	2457713	&	2458093	&	{\it CV}	&	65	\\
ZTF	&	[9]; Bellm et al. (2019)	&	2018 Mar 28	&	2458205	&	2459455	&	{\it zg, zr}	&	223	\\
		\hline
	\end{tabular}
	
\begin{flushleft}	
References:  [1] https://www.aavso.org/ , https://www.aavso.org/data-download    \newline
[2] http://crts.caltech.edu/ , http://nunuku.caltech.edu/cgi-bin/getcssconedb\_release\_img.cgi
[3] https://catalogs.mast.stsci.edu/panstarrs/     \newline
[4] https://noirlab.edu/science/programs/ctio/instruments/Dark-Energy-Camera , https://datalab.noirlab.edu/nscdr1/index.php      \newline
[5] https://archive.stsci.edu/missions-and-data/k2, https://archive.stsci.edu/k2/search\_retrieve.html
[6] https://fallingstar-data.com/forcedphot/      \newline
[7] https://catalina.lpl.arizona.edu/
[8] https://lasco-www.nrl.navy.mil/
[9] https://irsa.ipac.caltech.edu/cgi-bin/Gator/nph-scan?projshort=ZTF
\end{flushleft}

\end{table*}

\subsection{Photometric data}

Many observers and many surveys have followed the brightness of U Sco after it returned to a brightness level consistent with normal quiescent levels by 2010 March 28.  Table 1 contains a summary of the useable data collected, along with references and links.

The American Association of Variable Star Observers (AAVSO) reports magnitudes in the AAVSO International Database, which collects photometry from the best amateur observers worldwide.  These data are now all of professional quality with CCD photometers and modern well-calibrated comparison stars.  This collection is from observers with many affiliations, for example A. Oksanen (with observer code OAR) is affiliated with the Ursa Astronomical Society centred in Finland.  For others providing eclipse time series, W. Cooney (observer code CWT) is affiliated with the AAVSO, and G. Myers (observer code MGW) is the Past President of the AAVSO.  Notably, the activities of many of the U Sco observers are part of the very large and directed observing programs of the Center for Backyard Astrophysics (CBA, led by J. Patterson), with this group now providing the majority of the world's professional-quality photometry of CVs of all types.  Oksanen, Cooney, and Myers are all tireless CBA observers, and they have provided critical and voluminous light curve data for a variety of my previous research programs and papers.

Asteroid Terrestrial-impact Last Alert System (ATLAS) is a fully robotic 0.5-m f/2 Wright Schmidt telescope at Haleakala Observatory on the island of Maui, designed to detect small asteroids passing close to Earth.  The nominal limiting magnitude is $r$=18.0, with U Sco only marginal for detection.  The real scatter in each color has an RMS of 0.36 mag, and no eclipse can be picked out.  The use of non-standard filters (`cyan' and `orange') means that these magnitudes cannot be confidently combined with standard magnitudes from other sources for use in a long term light curve.

The Catalina Sky Survey (CSS) operates a sky survey with the primary purpose of seeking near-Earth asteroids.  The U Sco images came from two survey telescopes, a 0.7-m Schmidt telescope on Mount Bigelow, and a 1.5-m telescope on Mount Lemmon both near Tucson Arizona.  The CSS images showing U Sco were kindly passed along by E. Christensen (CSS PI).  I extracted the magnitudes from the images with the usual aperture photometry programs of \texttt{IRAF}.  The CCD images were taken with no filter, but the magnitudes were estimated for comparison stars with $V$-band magnitudes, resulting in a ``{\it CV}'' magnitude, which is close to the classic $V$-band yet with hard-to-determine relatively-small color corrections.  

The Catalina Real-Time Transient Survey (CRTS) uses images from the CSS to hunt for transients, compile light curves from the images, and make their photometry publicly available.  The whole on-line light curve for U Sco covers 2005--2013, but here, I will only be using the magnitudes from 2010.4 to 2013.5.

The observations at Cerro Tololo Inter-American Observatory (CTIO or CT) were made by myself with the 1.3-m telescope on Cerro Tololo in northern Chile.  These were service observing runs, designed to get time series centred on the primary eclipses, for purposes of contributing to the $O-C$ curve.  The comparison star magnitudes are given in Schaefer (2010), all standard analysis and programs (including \texttt{IRAF}) were used,  and these light curves are just a continuation of a very long series started in 1989 (Schaefer 1990; 2010; Schaefer \& Ringwald 1995, Schaefer et al. 2011).

Dark Energy Camera (DECam) is a wide field imaging array of CCDs mounted at the prime focus of the 4-m telescope at CTIO.  The only public domain data currently available are 24 images from the years 2013--2015.

After the failure of some gyroscopes on the {\it Kepler} spacecraft, it was repurposed into the K2 mission, where target fields scattered around the ecliptic were steadily observed for nearly 78 days.  U Sco was covered in Campaign 2, I proposed for short cadence observations of U Sco, and this was accepted (GO2007, PI Schaefer).  K2 provided nearly continuous photometry for 78 days from JD 2456894 (2014 August 24) to 2456972 (2014 November 10).  I have extracted the full light curve with standard analysis, resulting in 106,779 flux measures.  The one-sigma error bars were calculated by the normal pipeline programs of K2 for the Poisson noise and the standard corrections.  The resulting light curve shows no deviations or correlations with times of thruster firings by the {\it Kepler} spacecraft, nor with the exact positions of the image centroid, nor with the background, and it appears to be fully corrected for the known instrumental effects and any other artefacts.  The result is a wonderful 78 day light curve with 106,779 fluxes for a time resolution of 58.8 seconds.

Large Angle and Spectrometric Coronagraph Experiment (LASCO) is a set of three coronagraphs aboard the {\it SOHO} spacecraft, aimed directly at the Sun, with a field of view out to near 7.5$\degr$.  U Sco passes 3.6$\degr$ from the Sun every year on November 29 , and an eruption around the time of conjunction can only be visible with a space-borne coronagraph.  The relevant images from the C3 spectrograph were forwarded to me courtesy of K. Battams (US Naval Research Laboratory).  I flattened the field from the corona, identified the nearest and faintest stars, and observed that the position of U Sco has no point source.    C3 records white light, and the comparison star magnitudes are in the $V$-band, resulting in $CV$ magnitudes.  Stars as faint as $V$=9.33 are visible, but the 50 per cent detection threshold is near $V$=8.8.  These images were taken at half-day intervals.  The LASCO limits prove that U Sco did not erupt in the 16-day interval centred on the critical solar conjunction in late-2016.

Pan-STARRS is a sky survey covering $3\pi$ steradians of sky north of declination $-$30$\degr$ in five filters ({\it grizy}) from a 1.8-m telescope at Haleakala Observatory on the island of Maui in Hawaii.  For single epoch photometry, Pan-STARRS gets down to $g$=22.0 at five-sigma.  The available data are currently only the light curve from 2010--2014.

The Zwicky Transient Facility (ZTF) surveys the entire sky north of $-$29$\degr$ declination through two filter ($zg$ and $zr$) with the 48-inch Schmidt telescope at Palomar Observatory in California.  The limiting $zr$ magnitude is roughly 20.6.

All of the photometry (other than the K2 light curve) is presented in Table 2.  The full table, with 2920 lines of magnitudes, is available on-line as Supplementary Material, with the print version only showing the first and last five lines as examples.  The first column is the heliocentric Julian date (HJD) of the middle of the exposure, while the second column is the year.  The lines are all given in time order.  The third column gives the band or filter for the magnitude (with its one-sigma measurement error bar) in the next column.  The last column gives the source for the magnitude, keyed to Table 1.  The AAVSO observations are listed with the observer codes in parentheses.

\begin{table}
	\centering
	\caption{U Sco light curve after 2010.4, full table of 2920 magnitudes is in the on-line supplementary material}
	\begin{tabular}{lllll} 
		\hline
		Time (HJD)  &  Year  &  Band   &  Magnitude  &  Source  \\
		\hline
2455349.35974	&	2010.4164	&	CR	&	17.5	$\pm$	0.2	&	AAVSO(MLF)	\\
2455351.31173	&	2010.4218	&	CR	&	17.8	$\pm$	0.2	&	AAVSO(MLF)	\\
2455352.39171	&	2010.4247	&	CR	&	18.0	$\pm$	0.3	&	AAVSO(MLF)	\\
2455352.77254	&	2010.4258	&	V	&	17.19	$\pm$	0.08	&	CRTS	\\
2455352.78264	&	2010.4258	&	V	&	17.23	$\pm$	0.08	&	CRTS	\\
...	&		&		&				&		\\
2459655.27456	&	2022.2054	&	CV	&	18.759	$\pm$	0.022	&	AAVSO(MGW)	\\
2459655.28155	&	2022.2054	&	CV	&	18.627	$\pm$	0.016	&	AAVSO(MGW)	\\
2459655.28853	&	2022.2054	&	CV	&	18.523	$\pm$	0.014	&	AAVSO(MGW)	\\
2459655.29699	&	2022.2055	&	CV	&	18.433	$\pm$	0.016	&	AAVSO(MGW)	\\
2459655.30398	&	2022.2055	&	CV	&	18.422	$\pm$	0.031	&	AAVSO(MGW)	\\
		\hline
	\end{tabular}
\end{table}

The K2 light curve, with 106,779 fluxes is too long to conveniently fit into Table 2.  Instead, I have put the 58.8-s resolution light curve into Table 3.  All 106,779 fluxes are listed in the on-line Supplementary Material version, whereas the printed table only shows the first and last five lines as examples.  The times are in Barycentric Julian Date (BJD), which for our purposes is close to the HJD times used for the ground-based observations.  The fluxes and their one-sigma measurement errors are in units of count/second.

\begin{table}
	\centering
	\caption{K2 light curve for U Sco with 58.8 second resolution, full table of 106779 fluxes is in the on-line supplementary material}
	\begin{tabular}{ll} 
		\hline
		Time (BJD)  &  Flux (ct/sec)  \\
		\hline
2456894.800492	&	1180	$\pm$	27	\\
2456894.801174	&	1116	$\pm$	27	\\
2456894.801855	&	1228	$\pm$	27	\\
2456894.802536	&	1228	$\pm$	27	\\
2456894.803217	&	1218	$\pm$	27	\\
...	&				\\
2456972.055697	&	1107	$\pm$	31	\\
2456972.056378	&	1194	$\pm$	31	\\
2456972.057059	&	1188	$\pm$	31	\\
2456972.057740	&	1224	$\pm$	31	\\
2456972.058422	&	1138	$\pm$	31	\\
		\hline
	\end{tabular}
\end{table}

\subsection{Eclipse times}

The principle line of work in this paper is to measure the period changes in U Sco.  This task is to extract eclipse times, then calculate from the $O-C$ diagram the steady period change between eruptions ($\dot{P}$) and the sudden period change across eruptions ($\Delta P_{nova}$).

After extensive study with {\it Kepler} and {\it TESS} light curves, Schaefer (2021) concluded that the best eclipse times (i.e., with the smallest jitter from a smooth ephemeris) are for fitting parabolas to the minimum or to use bisectors of the light curve that involve only the lowest 50--80 per cent of the depth of the minimum.  

The eclipse time from the ZTF data was made with light curves sparsely sampled in time, but adequately sampled around the eclipse in phase space.  For these, the parabola fits were performed in phase space, with the period being known well enough.

I am adding 100 new U Sco eclipse times.  A total of 39 of the new eclipse times are from the ground-based light curves (Table 2), while the other 59 are from my K2 light curve (Table 3).  Further, I have added two eclipse times from 2003 made with the MDM 1.3-m telescope on Kitt Peak by J. Kemp (now at Middlebury College).  These 100 new eclipse times are presented in Table 4.  Table 4 also has added in the 67 eclipse times from before 2010.4, as collected in Schaefer (2011).  Many of the prior  eclipse times appear only in the on-line Supplementary Material, and they have already been listed in Schaefer (2011).  I have a grand total of 167 eclipse times for U Sco, from 1945 to 2022, spanning tightly across three eruptions.  A total of 121 eclipse times are my own personal data.

Table 4 has each of the eclipse times identified by a running integer ($N$) counting the number of orbits from some fiducial eclipse at the time of the start of the 1999 nova event.  For purposes of calculating the $O-C$ diagram, the predicted times of eclipse are calculated from some fiducial ephemeris, for which I will follow Schaefer (2011) and use an epoch of HJD 2451234.5387 and a period of 1.23054695 days.  The `observed-minus-calculated' time difference is between the observed time for each mid-eclipse and its time from the fiducial linear ephemeris.  These $O-C$ values for each observed eclipse are listed in the last column of Table 4.

\begin{table}
	\centering
	\caption{U Sco eclipse times 1945--2022, all 167 lines in supplementary material}
	\begin{tabular}{lllrr} 
		\hline
		UT Date  &  Telescope  &  $T_{obs}$ (HJD)  &  $N$  & $O-C$  \\
		\hline
1945 Jul 2 	&	Harvard	&	2431639.3000	$\pm$	0.0090	&	-15924	&	-0.0091	\\
1989 July 10	&	CT 0.9m	&	2447717.6064	$\pm$	0.0062	&	-2858	&	-0.0291	\\
1989 July 11	&	CT 0.9m	&	2447718.8481	$\pm$	0.0084	&	-2857	&	-0.0180	\\
1989 July 15	&	CT 0.9m	&	2447722.5406	$\pm$	0.0018	&	-2854	&	-0.0171	\\
1989 July 16	&	CT 0.9m	&	2447723.7675	$\pm$	0.0030	&	-2853	&	-0.0208	\\
...	&	&	&		&		\\
2003 Jun 4	&	MDM	&	2452794.8720	$\pm$	0.0008	&	1268	&	-0.0002	\\
2003 Jun 25	&	MDM	&	2452815.7849	$\pm$	0.0008	&	1285	&	-0.0066	\\
...	&	&	&		&		\\
2010 Mar 12 	&	Stockdale	&	2455268.2625	$\pm$	0.0020	&	3278	&	-0.0091	\\
2010 Mar 15 	&	Oksanen	&	2455270.7446	$\pm$	0.0009	&	3280	&	0.0119	\\
2010 Mar 16 	&	Krajci	&	2455271.9637	$\pm$	0.0031	&	3281	&	0.0005	\\
2010 Mar 26 	&	Oksanen	&	2455281.8158	$\pm$	0.0012	&	3289	&	0.0082	\\
2010 Mar 31 	&	Oksanen	&	2455286.7411	$\pm$	0.0025	&	3293	&	0.0113	\\
2010 May 18	&	CT 0.9m	&	2455334.7211	$\pm$	0.0009	&	3332	&	0.0000	\\
2010 Jun 29	&	CT 0.9m	&	2455376.5650	$\pm$	0.0035	&	3366	&	0.0053	\\
2010 Jul 5	&	MDM	&	2455382.7126	$\pm$	0.0008	&	3371	&	0.0001	\\
2010 Jul 10	&	CT 0.9m	&	2455387.6395	$\pm$	0.0010	&	3375	&	0.0048	\\
2010 Aug 15	&	Oksanen	&	2455424.5565	$\pm$	0.0010	&	3405	&	0.0054	\\
2011 Apr 14	&	CT 0.9m	&	2455665.7492	$\pm$	0.0009	&	3601	&	0.0109	\\
2011 May 5	&	CT 0.9m	&	2455686.6672	$\pm$	0.0016	&	3618	&	0.0096	\\
2011 May 27	&	CT 0.9m	&	2455708.8169	$\pm$	0.0015	&	3636	&	0.0095	\\
2011 June 11	&	CT 0.9m	&	2455723.5826	$\pm$	0.0010	&	3648	&	0.0086	\\
2011 June 27	&	CT 0.9m	&	2455739.5762	$\pm$	0.0012	&	3661	&	0.0051	\\
2011 July 23	&	CT 0.9m	&	2455766.6569	$\pm$	0.0009	&	3683	&	0.0138	\\
2012 Mar 26	&	CT 0.9m	&	2456012.7698	$\pm$	0.0007	&	3883	&	0.0173	\\
2012 May 2	&	CT 0.9m	&	2456049.6833	$\pm$	0.0010	&	3913	&	0.0144	\\
2012 May 28	&	CT 0.9m	&	2456076.7509	$\pm$	0.0008	&	3935	&	0.0100	\\
2012 July 25	&	CT 0.9m	&	2456134.5890	$\pm$	0.0009	&	3982	&	0.0123	\\
2012 Aug 10	&	CT 0.9m	&	2456150.5850	$\pm$	0.0010	&	3995	&	0.0112	\\
2014 Aug 25	&	K2	&	2456895.0641	$\pm$	0.0010	&	4600	&	0.0094	\\
2014 Aug 26	&	K2	&	2456896.2991	$\pm$	0.0009	&	4601	&	0.0139	\\
2014 Aug 29	&	K2	&	2456898.7582	$\pm$	0.0009	&	4603	&	0.0119	\\
2014 Aug 30	&	K2	&	2456899.9905	$\pm$	0.0007	&	4604	&	0.0136	\\
2014 Aug 31	&	K2	&	2456901.2212	$\pm$	0.0007	&	4605	&	0.0138	\\
2014 Sep 1	&	K2	&	2456902.4519	$\pm$	0.0009	&	4606	&	0.0139	\\
2014 Sep 3	&	K2	&	2456903.6797	$\pm$	0.0009	&	4607	&	0.0112	\\
2014 Sep 4	&	K2	&	2456904.9121	$\pm$	0.0009	&	4608	&	0.0131	\\
2014 Sep 5	&	K2	&	2456906.1439	$\pm$	0.0009	&	4609	&	0.0143	\\
2014 Sep 6	&	K2	&	2456907.3743	$\pm$	0.0007	&	4610	&	0.0142	\\
2014 Sep 8	&	K2	&	2456908.6017	$\pm$	0.0007	&	4611	&	0.0110	\\
2014 Sep 9	&	K2	&	2456909.8351	$\pm$	0.0009	&	4612	&	0.0139	\\
2014 Sep 10	&	K2	&	2456911.0670	$\pm$	0.0009	&	4613	&	0.0152	\\
2014 Sep 11	&	K2	&	2456912.2954	$\pm$	0.0010	&	4614	&	0.0131	\\
2014 Sep 13	&	K2	&	2456913.5182	$\pm$	0.0011	&	4615	&	0.0053	\\
2014 Sep 14	&	K2	&	2456914.7501	$\pm$	0.0011	&	4616	&	0.0067	\\
2014 Sep 15	&	K2	&	2456915.9801	$\pm$	0.0012	&	4617	&	0.0061	\\
2014 Sep 16	&	K2	&	2456917.2124	$\pm$	0.0017	&	4618	&	0.0079	\\
2014 Sep 17	&	K2	&	2456918.4442	$\pm$	0.0015	&	4619	&	0.0091	\\
2014 Sep 19	&	K2	&	2456919.6708	$\pm$	0.0017	&	4620	&	0.0052	\\
2014 Sep 20	&	K2	&	2456920.9031	$\pm$	0.0012	&	4621	&	0.0069	\\
2014 Sep 21	&	K2	&	2456922.1363	$\pm$	0.0015	&	4622	&	0.0096	\\
2014 Sep 22	&	K2	&	2456923.3678	$\pm$	0.0012	&	4623	&	0.0106	\\
2014 Sep 25	&	K2	&	2456925.8276	$\pm$	0.0012	&	4625	&	0.0093	\\
2014 Sep 26	&	K2	&	2456927.0565	$\pm$	0.0015	&	4626	&	0.0076	\\
2014 Sep 27	&	K2	&	2456928.2893	$\pm$	0.0014	&	4627	&	0.0099	\\
2014 Sep 29	&	K2	&	2456929.5245	$\pm$	0.0014	&	4628	&	0.0145	\\
2014 Sep 29	&	K2	&	2456930.7498	$\pm$	0.0013	&	4629	&	0.0093	\\
2014 Oct 1	&	K2	&	2456931.9823	$\pm$	0.0013	&	4630	&	0.0112	\\
2014 Oct 2	&	K2	&	2456933.2111	$\pm$	0.0014	&	4631	&	0.0095	\\
2014 Oct 5	&	K2	&	2456935.6742	$\pm$	0.0021	&	4633	&	0.0115	\\
2014 Oct 6	&	K2	&	2456936.9083	$\pm$	0.0013	&	4634	&	0.0150	\\
2014 Oct 7	&	K2	&	2456938.1387	$\pm$	0.0014	&	4635	&	0.0149	\\

		\hline
	\end{tabular}
\end{table}

\begin{table}
	\centering
	\contcaption{U Sco eclipse times 1945--2022, all 167 lines in supplementary material}
	\label{tab:continued}
	\begin{tabular}{lllrr}
		\hline
		UT Date  &  Telescope  &  $T_{obs}$ (HJD)  &  $N$  & $O-C$  \\
		\hline

2014 Oct 8	&	K2	&	2456939.3644	$\pm$	0.0010	&	4636	&	0.0100	\\
2014 Oct 10	&	K2	&	2456940.6075	$\pm$	0.0010	&	4637	&	0.0226	\\
2014 Oct 11	&	K2	&	2456941.8305	$\pm$	0.0009	&	4638	&	0.0150	\\
2014 Oct 12	&	K2	&	2456943.0654	$\pm$	0.0007	&	4639	&	0.0194	\\
2014 Oct 13	&	K2	&	2456944.2902	$\pm$	0.0009	&	4640	&	0.0137	\\
2014 Oct 15	&	K2	&	2456945.5247	$\pm$	0.0016	&	4641	&	0.0176	\\
2014 Oct 16	&	K2	&	2456946.7497	$\pm$	0.0009	&	4642	&	0.0121	\\
2014 Oct 17	&	K2	&	2456947.9825	$\pm$	0.0007	&	4643	&	0.0143	\\
2014 Oct 18	&	K2	&	2456949.2108	$\pm$	0.0011	&	4644	&	0.0121	\\
2014 Oct 19	&	K2	&	2456950.4447	$\pm$	0.0010	&	4645	&	0.0154	\\
2014 Oct 22	&	K2	&	2456952.9058	$\pm$	0.0009	&	4647	&	0.0154	\\
2014 Oct 23	&	K2	&	2456954.1270	$\pm$	0.0011	&	4648	&	0.0061	\\
2014 Oct 24	&	K2	&	2456955.3605	$\pm$	0.0012	&	4649	&	0.0090	\\
2014 Oct 26	&	K2	&	2456956.5928	$\pm$	0.0014	&	4650	&	0.0108	\\
2014 Oct 27	&	K2	&	2456957.8256	$\pm$	0.0013	&	4651	&	0.0130	\\
2014 Oct 28	&	K2	&	2456959.0584	$\pm$	0.0013	&	4652	&	0.0153	\\
2014 Oct 29	&	K2	&	2456960.2862	$\pm$	0.0012	&	4653	&	0.0125	\\
2014 Oct 31	&	K2	&	2456961.5165	$\pm$	0.0012	&	4654	&	0.0123	\\
2014 Nov 1	&	K2	&	2456962.7431	$\pm$	0.0015	&	4655	&	0.0083	\\
2014 Nov 2	&	K2	&	2456963.9825	$\pm$	0.0012	&	4656	&	0.0172	\\
2014 Nov 3	&	K2	&	2456965.2100	$\pm$	0.0010	&	4657	&	0.0142	\\
2014 Nov 4	&	K2	&	2456966.4425	$\pm$	0.0010	&	4658	&	0.0161	\\
2014 Nov 6	&	K2	&	2456967.6703	$\pm$	0.0010	&	4659	&	0.0134	\\
2014 Nov 7	&	K2	&	2456968.8950	$\pm$	0.0011	&	4660	&	0.0075	\\
2014 Nov 8	&	K2	&	2456970.1334	$\pm$	0.0008	&	4661	&	0.0154	\\
2014 Nov 9	&	K2	&	2456971.3649	$\pm$	0.0010	&	4662	&	0.0163	\\
2016 July 22	&	CT 1.3-m	&	2457591.5515	$\pm$	0.0006	&	5166	&	0.0073	\\
2016 Aug 6	&	CT 1.3-m	&	2457607.5433	$\pm$	0.0007	&	5179	&	0.0019	\\
2017 May 19	&	Pan	&	2457893.0323	$\pm$	0.0011	&	5411	&	0.0041	\\
2019 Apr 19	&	G. Myers	&	2458593.2114	$\pm$	0.0031	&	5980	&	0.0019	\\
2019 May 15	&	G. Myers	&	2458619.0647	$\pm$	0.0011	&	6001	&	0.0138	\\
2019 May 21	&	G. Myers	&	2458625.2146	$\pm$	0.0028	&	6006	&	0.0109	\\
2019 May 26	&	G. Myers	&	2458630.1436	$\pm$	0.0007	&	6010	&	0.0177	\\
2019 June 5	&	G. Myers	&	2458639.9845	$\pm$	0.0027	&	6018	&	0.0143	\\
2019 June 21	&	G. Myers	&	2458655.9785	$\pm$	0.0014	&	6031	&	0.0111	\\
2019 July 2	&	G. Myers	&	2458667.0601	$\pm$	0.0009	&	6040	&	0.0178	\\
2019 July 11	&	W. Cooney	&	2458675.6711	$\pm$	0.0022	&	6047	&	0.0150	\\
2020 Mar 20	&	G. Myers	&	2458929.1638	$\pm$	0.0016	&	6253	&	0.0150	\\
2020 April 5	&	G. Myers	&	2458945.1620	$\pm$	0.0010	&	6266	&	0.0161	\\
2020 April 16	&	G. Myers	&	2458956.2361	$\pm$	0.0011	&	6275	&	0.0153	\\
2020 May 18	&	G. Myers	&	2458988.2365	$\pm$	0.0014	&	6301	&	0.0215	\\
2020 June 12	&	W. Cooney	&	2459012.8399	$\pm$	0.0018	&	6321	&	0.0139	\\
2020 June 29	&	G. Myers	&	2459030.0747	$\pm$	0.0012	&	6335	&	0.0211	\\
2020 Aug 5	&	G. Myers	&	2459066.9823	$\pm$	0.0026	&	6365	&	0.0123	\\
2020 Sep 11	&	G. Myers	&	2459103.9030	$\pm$	0.0029	&	6395	&	0.0166	\\
2021 April 14	&	G. Myers	&	2459319.2452	$\pm$	0.0014	&	6570	&	0.0130	\\
2021 April 19	&	G. Myers	&	2459324.1680	$\pm$	0.0010	&	6574	&	0.0137	\\
2021 April 24	&	G. Myers	&	2459329.0971	$\pm$	0.0017	&	6578	&	0.0206	\\
2021 May 9	&	ZTF	&	2459343.8735	$\pm$	0.0045	&	6590	&	0.0304	\\
2021 May 31	&	G. Myers	&	2459366.0111	$\pm$	0.0010	&	6608	&	0.0182	\\
2021 July 7	&	G. Myers	&	2459402.9262	$\pm$	0.0016	&	6638	&	0.0168	\\
2021 July 29	&	G. Myers	&	2459425.0707	$\pm$	0.0005	&	6656	&	0.0115	\\
2021 Aug 19	&	G. Myers	&	2459446.0001	$\pm$	0.0023	&	6673	&	0.0216	\\
2022 Mar 16	&	G. Myers	&	2459655.1917	$\pm$	0.0011	&	6843	&	0.0202	\\
		\hline
	\end{tabular}
\end{table}

\subsection{Phased light curve}

The phased light curves show the primary eclipses, secondary eclipses, and any asymmetries arising from the beaming pattern of the disc.  The 1988--1989 phased light curve in the $B$ band was shown in fig. 3 of Schaefer (1990), which was the discovery of the U Sco eclipses.  Phased light curves in $B$, $V$, and $I$ for 2000--2010 are in fig. 3 of Schaefer et al. (2010) plus figures 45 and 47 of Schaefer (2010).  In this section, I report on the phased light curve for the light curve in quiescence after the 2010 eruption has ended.

The post-2010 magnitudes in quiescence (see Table 2) come in a wide variety of bands.  The light curves for each band are all similar to each other, and any one band makes for a moderately sparse phased plot.  An adequate solution to get a good plot is to normalize all the various bands to each other to create one plot with good coverage from many measures.  The color variations throughout the cycle are moderate, hence the scatter added by this procedure is negligibly small.  A larger source of scatter is the over-all rise and fall of the average over months and years, as well as the fast flickering.  Still, the folded light curve is adequate to show the details of the eclipses, the quadratures, and any asymmetries outside of eclipse.  The magnitudes are all normalized to the $V$-band.  To normalize the light curves, I have taken pairs of magnitudes in different bands taken nearly simultaneously, with the average differences making the corrections to $V$-band.  The correction from $B$ to $V$ (i.e., $\langle B-V \rangle$) is 0.54 mag.  The correction for $zg$ is $+$0.36 (i.e., $\langle$$zg$$-$$V$$\rangle$=$+$0.36), for $CV$ is $+$0.06, for $zr$ is $-$0.10, for $R$ is $-$0.34, and for $I$ is $-$0.80.  I do not have adequate information to calculate confident offsets for the $g$, $visual$, $CR$, `cyan', `orange', $i$, $z$, $y$, and Kepler bands.

The post-2010 phased light curve is made from the normalized magnitudes from Table 2, with the $O-C$ ephemeris (see equation 4).  This plot is shown in Fig. 1.

\begin{figure}
	\includegraphics[width=1.05\columnwidth]{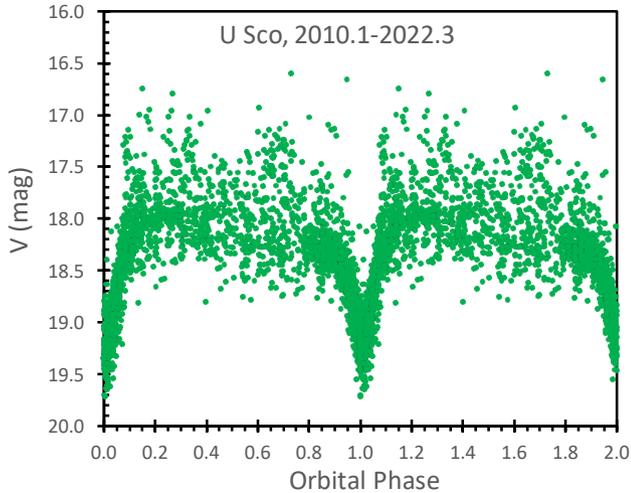}
    \caption{The phased and folded light curve for 2010.1 to 2022.3.  This displays the prominent primary eclipse, no apparent secondary eclipse, and a definitely asymmetry in the levels for the quadratures around phases 0.25 and 0.75.  The asymmetry between the quadratures is around 0.2 mag, similar to that seen in the folded K2 light curve.  The scatter in the light curve is mainly cause by the usual flickering plus the rising and falling levels across the months and years.  The flickering is significantly smaller during the primary eclipse.  The orbital phase is calculated with a period $P$=1.23054695 days and epoch $E_0$ at HJD 2451234.5387, for which the real mid-eclipse times occur for phases close to $+$0.01 days.  The plotted magnitudes for each band have been normalized to the level of the $V$-band.  Each point is plotted twice, once for the normal phase range 0--1, and again for phase range 1--2.}
\end{figure}

The light curve shows an asymmetry in the out-of-eclipse levels, in that the light curve around the 0.25 phase quadrature is roughly 0.20 mag brighter than the level around the 0.75 phase quadrature.   This asymmetry is produced by the differing brightness emitted by the system as viewed from differing directions as the binary rotates, for which I speculate that the extra light around phase 0.25 is from irradiation on the inside of the raised hotspot, where the accretion stream hits the accretion disc.  This asymmetry is similar for the various filters from $B$ to $I$.  The K2 light curve (see Section 3.1) shows a similar asymmetry with an amplitude of 0.12 mag.

This asymmetry is not present in the 1988-1989 $B$-band light curve (Schaefer 1990).  Nor does the asymmetry appear in the $B$ and $V$ folded light curves for 2000--2010 (Schaefer et al. 2010).  Something happened across the 2010 eruption.  Perhaps the relative size of the hotspot was smaller before 2010 than after.

\subsection{Long-term variations}

The brightness level of U Sco varies on all time-scales, from minutes to years.  My analysis of the U Sco spectral energy distribution shows that $\sim$80 per cent of the $V$-band light is from the accretion disc\footnote{This analysis is described in Schaefer (2022), while using an SED taken from $Galex$, Pan-STARRS, 2MASS, $WISE$, plus my own UBVRIJHK measures from Cerro Tololo (Schaefer 2010), with $E(B-V)$=0.31 mag.}, hence the light curve is predominantly a measure of the accretion rate.  The average long-term brightness level tracks the mean accretion rate, and the time-integrated brightness level tracks the mass accumulated on the surface of the white dwarf between eruptions.

I have constructed an averaged long-term light curve for U Sco in quiescence and outside eclipse from 1976 to 2022.  After collecting the magnitudes, I have normalized them to the $V$-band.  The magnitudes are then averaged in time.  The first averaging was to construct nightly averages.  The reason is that a single night with many observations from a time series would be given a very high weight when compared to many nights with only one measured magnitude, and such would not provide a realistic measure of the long-term brightness level of U Sco.  With these nightly averages, the highly irregular cadence can produce biased averages on longer time-scales.  In particular, some years have many observations and some have few.  If all the nightly measures are simply averaged together between an eruption, the resulting brightness level will be nearly that of the one year with the most observations, rather than the brightness level throughout the inter-eruption interval.  A good solution is to form the nightly means into averages over half-year intervals, and then use these half-year averages for determining the variations in the mean accretion rate.  These averages are presented in Fig. 2.

\begin{figure}
	\includegraphics[width=1.05\columnwidth]{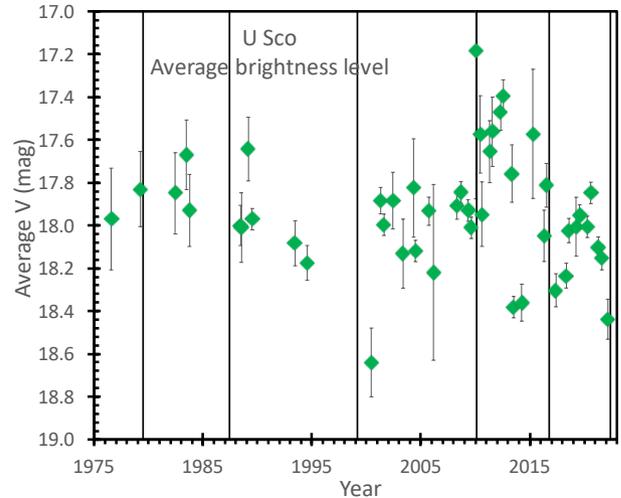}
    \caption{Long-term averaged light curve for U Sco from 1976 to 2022.  The brightness level and the accretion rate varies on all time-scales from minutes to years, all as part of a power law in the power-spectrum-distribution.  This plot shows the light curve in quiescence, normalized to the $V$-band, and averaged to roughly half-year time bins.  The vertical lines indicate the dates of the U Sco eruptions.  With this, the rise and fall of the accretion between eruptions is visible, with the mass accreted between eruptions accumulating up until the trigger mass is collected to start the next eruption.  Importantly, the brightness (and hence the accretion rate) was very high from 2010.1 to 2016.8, and this explains why the eruption occurred so soon after the 2010.1 eruption.  Further, the plot shows the accumulation rate of mass from 2016.8 to 2022.3 to be somewhat below average, pointing to a somewhat longer-than-average $\delta$Y, contrary to observations.}
\end{figure}

The uncertainties for the half-yearly averages in Fig. 2 have a variety of problems, where a formal propagation of the reported measurement errors does not produce useful results.  One problem is that the formal measurement errors, as always tabulated, do not include the random jitter from ordinary flickering, which act like an additional random instance-to-instance noise source.  To represent these sources of noise, I have added in quadrature the reported measurement error with 0.08 mag for $B$ and $V$, or with 0.04 mag for the redder filters.  A further problem is the uncertainty in the conversion to $V$-band, where the many observers' adaptation of differing filters and CCD spectral sensitivities make for variations in the corrections to $V$-band.  These color corrections are now impossible to recover, so I can only increase the associated uncertainties.  To represent this, the error bars for the individual magnitudes have the further addition in quadrature of 0.10 mag for most bands other than $V$, or 0.2 mag for $I$ and $CV$ bands.  For nights with more than one measure, the nightly average was formed with error bars taken to be the largest value of the RMS divided by the square root of the number of input magnitudes, the average error divided by the square root of the number of input magnitudes, and 0.05 mag.  This same procedure was used to produce the error bars for the half-yearly averages from the nightly averages.  This method does not account for the short-term correlations between magnitudes taken over a short number of nights (like in one observing run), with these correlations being poorly known and producing a smaller scatter than might be appropriate over the entire half-year interval.  This complicated means of estimating the real error bars for the half-yearly averages cannot claim to have any high accuracy, mainly because of unknown color corrections, night-to-night correlations, and poor sampling.  Perhaps a better way to see the effective uncertainties is simply to look at the RMS scatter of the points in Fig. 2, for which $\pm$0.20 mag might be the best overall estimate of the typical error bars.  

Fig. 2 shows some apparently significant trends over the years.  A critical point is that U Sco was substantially brighter than average over the 2010.1 to 2016.8 interval.  Then the accretion rate was particularly high, and the required trigger mass was accumulated much faster than the usual 10-years.  This explains why U Sco erupted in 2016.8, only 6.7 years after its previous eruption.  Another apparently significant trend is that U Sco has a rise and fall in brightness by roughly a third of a magnitude from 2017 to 2022.  The average level during this interval is somewhat fainter than the overall average level, indicating a lower than average level of accretion.  With this trend, it is hard to explain why U Sco had its shortest $\delta$Y of 5.6 years despite having a lower than average accretion rate.

\subsection{Time interval between eruptions}

The time interval between eruptions of RNe can be explained and predicted from the quiescent brightness levels in the preceding inter-eruption interval (Schaefer 2005).  The idea is to use the observed brightness as a measure proportional to the mass accretion rate, integrate the rate since the prior eruption to show the accumulating mass on the surface of the white dwarf, and predict when the accumulated mass reaches the trigger mass.  The trigger mass and scaling factors are calibrated from prior time intervals for that particular RN.   

This method starts with the idea that the white dwarf needs to accumulate some critical threshold of mass ($M_{trigger}$) to trigger the nova event.  This trigger mass depends primarily on the white dwarf mass and the composition of the in-falling material, with these conditions remaining constant from eruption to eruption.  The question about the time between two eruptions is to know when $M_{trigger}$ has been accumulated.  For notation, the year of a nova peak ($Y$) is followed by an interval of $\delta Y$ until the next eruption, and the accretion rate is denoted $\dot{M}$.  The equation for the trigger mass as an integration of $\dot{M}$ over time $t$ is
\begin{equation}
M_{trigger} = \int_{Y}^{Y+\delta Y} \dot{M} dt.
\end{equation}
For use in equation (1), the accretion rate is a function of the nova brightness.  The optical flux can be defined as $F$=10$^{-0.4(V-20)}$.  Here, the unit of flux is that from a $V$=20 star.    With U Sco having a long orbital period, the optical flux will be in the usual $F_{\nu} \propto \nu ^{1/3}$ region of the spectral energy distribution, for which $F$$\propto$$\dot{M}^{2/3}$, or $\dot{M}$=C$F^{1.5}$.  Here, $C$ is a constant that varies from nova-to-nova, is hard to know with any useful accuracy, and has the arcane units of solar-masses per year per $F^{1.5}$.  I have confirmed this relation for the explicit conditions of U Sco by means of a direct integration with the standard $\alpha$-disc model. For simplicity and to a useful approximation, the integral in equation (1) can be evaluated for some average $V$-magnitude over the entire inter-eruption interval, $\langle V \rangle$.  The time between eruptions is 
\begin{equation}
\delta Y = (M_{trigger}  / C)~ 10^{0.6(\langle V \rangle - 20)} .
\end{equation}
$M_{trigger}$ is hard to know with useful accuracy, but is a constant from eruption-to-eruption for a recurrent nova.  The constant $M_{trigger}/C$ should be a constant from eruption-to-eruption on a given nova, and can best be evaluated from each inter-eruption interval.

\begin{table}
	\centering
	\caption{Time between eruptions and the inter-eruption brightness level}
	\begin{tabular}{lllll} 
		\hline
		$Y$ Interval (years)  &  $\delta Y$    &   $N_{mags}$    &    $\langle V \rangle$    &   $M_{trigger}/C$ (Eq. 3)  \\
		\hline
1969.1 to 1979.5	&	10.4	&	2	&	17.90	&	150 	$\pm$	38	\\
1979.5 to 1987.4	&	7.9	&	4	&	17.82	&	130	$\pm$	38	\\
1987.4 to 1999.2	&	11.8	&	26	&	17.98	&	128	$\pm$	28	\\
1999.2 to 2010.1	&	10.9	&	221	&	17.97	&	124	$\pm$	8	\\
2010.1 to 2016.8	&	6.7	&	72	&	17.78	&	122	$\pm$	8	\\
2016.8 to 2022.4	&	5.6	&	329	&	18.11	&	59 $\pm$ 3	\\
		\hline
	\end{tabular}
\end{table}

The accretion rate is not linear in F, and it is somewhat more accurate to perform the integral in equation (1) as a summation over smaller time intervals, $\Delta t$.  A further improvement is to correct from the total brightness to the brightness of the disc alone.  From my fitting to the spectral energy distribution, I estimate that roughly 20 per cent of the $V$-band flux is from the companion, with the companion light being nearly a constant, and $F$-$F_{comp}$is the disc light that is related to the accretion rate.  The exact value of $F_{comp}$ is not known, but high accuracy is not critical, while a reasonable value is 1.0 (where $V$=20 is one unit of flux).  A better calculation from equation (1) is
\begin{equation}
M_{trigger} / C = \sum [10^{-0.4*(\langle V \rangle -20)}-1.0)]^{1.5} ~ \Delta t,
\end{equation}
where the summation is over all the intervals $\Delta t$ (each with average magnitude $\langle V \rangle$) between two successive eruptions.  The $M_{trigger}/C$ values can be calculated for each inter-eruption interval after 1969.  For this, I have used half-year intervals for $\Delta t$, like in Fig. 2.  For intervals with no input, I have interpolated from adjacent intervals, and assigned an error bar of $\pm$0.3 mag.  The result is six values of $M_{trigger}/C$, as listed in Table 5.

As discussed in Section 2.4, the uncertainties on the input magnitudes are poorly known.  The error bars from Fig. 2 have been propagated so as to produce error bars on the six $M_{trigger}/C$ values.  The first two time intervals have the expected large error bars, due to their small number of input magnitudes.  The last interval has a rather small error bar.  

The first five inter-eruption intervals are consistent with a constant value of near 125, while the last interval has a value that appears to be significantly lower at 59.  This leaves us with the conundrum of understanding this variation.  I see no reason to think that the last interval has any large systematic or observational error, and the error bars make it highly significant that the last interval has a value close to 2$\times$ smaller than in the previous intervals.  This observational result is contrary to the strong expectation that both $M_{trigger}$ and $C$ should be constant from eruption-to-eruption.  I do not know how to resolve this strong difference between observation and theory.

For the critical question of 2018, we need to recognize when the $M_{trigger}$ has accumulated such that the next eruption would occur.  For this question, only the first four inter-eruption intervals can be used, and these have a weighted mean value of 125$\pm$7.  For the summation in equation (3), $M_{trigger}/C$ reaches the trigger level in the year 2017.1$\pm$0.6.  My similar calculation in 2018 is what pointed to the eruption having already happened.

The 2$\times$ smaller value of $M_{trigger}/C$ for the last inter-eruption interval makes for a substantial uncertainty in knowing the effective value of $M_{trigger}/C$ for use in making further predictions.  If we apply this smaller value to the interval after the 2010 eruption, we would still have realized that the eruption had already occurred.  For knowing the true and unchanging value of $M_{trigger}/C$, if it is indeed unchanging, then taking a simple weighted mean of all six values would not be good, as the one far-outlier has the happenstance of having by-far the smallest error, and the weighted mean value of 72$\pm$2 is clearly a bad representation of the data.  A straight average of the six values  (119$\pm$13) is no better, while quoting a range (59--125) at least covers the cases.  With the variations from 59 to 125 being highly significant, likely the best answer is simply the realization that the trigger mass is not constant.  This would go along with the cases that $\Delta P$ and $\dot{P}$ are changing eruption-to-eruption, all this being invisible with no significant detected changes in the eruption optical photometry and spectroscopy.

\subsection{Limits on eruption around 2017.0$\pm$0.4}

The optical brightness after the 2010 eruption is pointing towards an eruption in 2017.1$\pm$0.6.  A kink in the $O-C$ curve proves that an eruption was missed in 2016.9$\pm$0.6.  (The date of this kink will be analyzed in Section 4.4.)  The missed eruption peaked at 2017.0$\pm$0.4.  Observed magnitudes after the 2015/2016 solar gap and before the 2017/2018 solar gaps prove that no eruption peaked inside the time interval 2015.9 to 2017.8, except for the 2016/2107 solar gap.  So the missed eruption occurred during the 2016/2017 solar gap.

Starting in 2018, I made a widespread search for earlier photometry to either discover U Sco in the tail of an eruption, or to at least constrain the possible dates.  Disappointingly, none of the collected photometry shows U Sco above its background level.  I can only constrain the date of the missed eruption as much as possible.  Table 2 gives all magnitudes and limits for use in defining each solar gap.

For each of the yearly solar gaps, I can tabulate the time intervals between the last observation of one season and the first observation of the next season.  For the 2010/2011 gap the interval was 105 days, while subsequent gaps are 103, 96, 98, 99, 101, 146, 80, 129, 86, 94, and 96 days up until the 2021/2022 gap.  The 2016/2017 and 2017/2018 gaps are complicated by the LASCO solar coronagraph limits around the times of solar conjunction.

The 2016/2017 seasonal gap started with the last ground-based observation of 2016 (on HJD 2457635.9) by A. Shaw (observer code SVA, an AAVSO member living in Australia) with a visual observation that U Sco was fainter than 13.1 magnitude.  Any eruption during this solar gap must have peaked on 2457637 or later.  The first ground-based observation of 2017 was on 2457782.2 by R. Stubbings (observer code SRX, affiliated with the Variable Star Section of the Royal Astronomical Society of New Zealand while living in Australia) with a visual observation that U Sco was fainter than 15.0 mag.  U Sco in eruption falls to this level at a time 28 days after the eruption peak, and Stubbings' limit shows that any peak must have been before 2457754.  A more constraining observation was on 2457790.6 by B. Monard (observer code MLF, affiliated with the CBA, observing from South Africa), where U Sco was detected at the level of $CR$=18.4$\pm$0.3.  With the nova at its quiescent level, the peak of any missed eruption could only be more than 60 days earlier, which is to say before 2457730.  These ground-based observation limit any possible nova peak to be in the 93 day interval from 2457637 to 2457730.  Further, the LASCO solar coronagraph limits that $CV$$>$8.8 at a half day cadence from 2457713.5 to 2457729.5.  U Sco is brighter than 8.8 mag only for a two day interval after the very fast rise to the nova peak, hence the missed eruption did not peak anytime in the interval 2457712 to 2457730.  Combined with the ground-based observations, any missed peak could only occur in the single interval 2457637 to 2457712.  This is a 75 day interval from 2016 September 5 to 2016 November 19, centred on 2016.78, with a maximal uncertainty of 0.10 years.

I looked far and wide for any useable images covering the years 2015--2019, in a vain hope to catch U Sco in eruption.  The useful sources have already been listed in Table 1.  With the lack of any measure of U Sco being above its quiescent level, this effort turned to putting limits on the possible dates of a missed eruption, principally by trying to push in on times when U Sco was close to the Sun.  It might be useful to briefly indicate which sources did not have any useful data for U Sco.  APASS, AAVSOnet, SMARTS, the NOAO archives, ASAS, and ASAS-SN did not have any useful coverage.  X-ray all-sky surveys (including {\it Swift} XRT) could only detect U Sco during eruption, but not when it was within 46$\degr$ of the Sun.  In hard X-ray and $\gamma$-rays, the various surveys (including {\it Fermi} GBM, {\it Swift} BAT, and MAXI) did not detect the 2010 eruption and could not detect any later missed eruption.  A number of perennial surveys did not have any coverage, including  LONEOS, NEAT, SMEI, Palomar Quest, LINEAR, STEREO, and Spacewatch.

In the end, from the light curve data, the only possible dates for a missed eruption is during a 75-day interval centred on 2016.78 (JD 2457674, 2016 October 12).

\section{{\it KEPLER} LIGHT CURVE}

The short cadence K2 light curve has a time resolution of 58.8 seconds, steadily, with only a few small gaps for the entire 78 days.  A sample five-day segment is shown in Fig. 3.  My full light curve, with 106779 fluxes, is presented in Table 3, with most of the data appearing only as on-line supplementary material.

\begin{figure}
	\includegraphics[width=1.1\columnwidth]{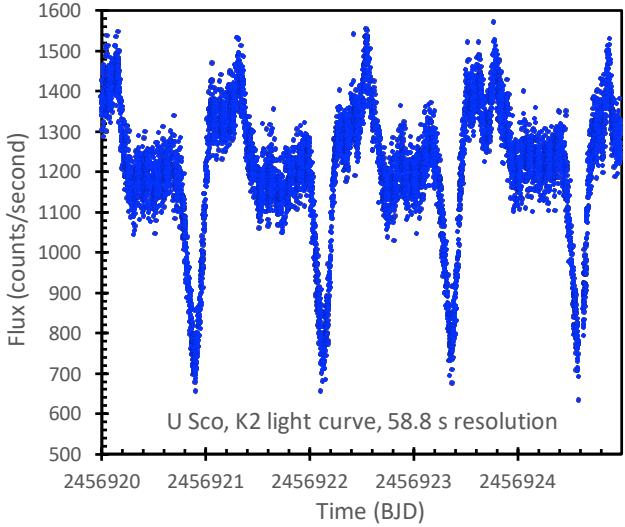}
    \caption{K2 light curve with 58.8 second resolution for a typical five-day interval, showing four deep primary eclipses.  The light curve shows a consistent drop soon before half phase, with this looking like the ingress for a secondary eclipse, with some sort of an asymmetry between the quadratures masking the egress.  The one-sigma photometric error bars are 27 count/second.  The observed vertical scatter in flux is intrinsic fast variability, i.e., flickering.}
\end{figure}

\subsection{Orbital variations}

We can see the orbital variations in the K2 light curve from looking at Fig. 3.  Still, it is useful to construct a folded and averaged light curve.  With all the K2 light curve, nearly 62 orbital cycles all averaged together.  (The number of orbits in the fold, 62, is different from the number of fitted eclipse times, 59, because 3 eclipses have small data gaps that would produce possibly-biased times if included.)  The resultant light curve is averaging over 62 instances of flickering.

I constructed the phased and averaged light curve from all the K2 data.  The phasing was shifted from the fiducial epoch for the $O-C$ curve to an epoch of HJD 2451234.5509 to place the average K2 eclipse times at zero phase.  Further, the daily ups-and-downs of the light curve were smoothly normalized to a flat light curve, to minimize the scatter.  (This analysis is presented in Section 3.2.  This normalization does not affect the shape of the phased light curve, but it does substantially reduce the scatter for the fluxes of the input fluxes inside each phase bin.)  The normalization was done with a multiplication factor, as appropriate for the flux levels changing with the accretion rate.  Once the fluxes were folded, they were divided into 1000 phase bins, with the average flux calculated for each bin.  The formal uncertainty is the RMS scatter of the flux in each bin divided by the square root of the number of input fluxes.  Even with a thousand phase bins, each phase bin has 107 fluxes.  The resultant phased and averaged light curve is presented in Fig. 4 and Table 6.  A close-up of the primary minimum is in Fig. 5.

\begin{figure}
	\includegraphics[width=1.07\columnwidth]{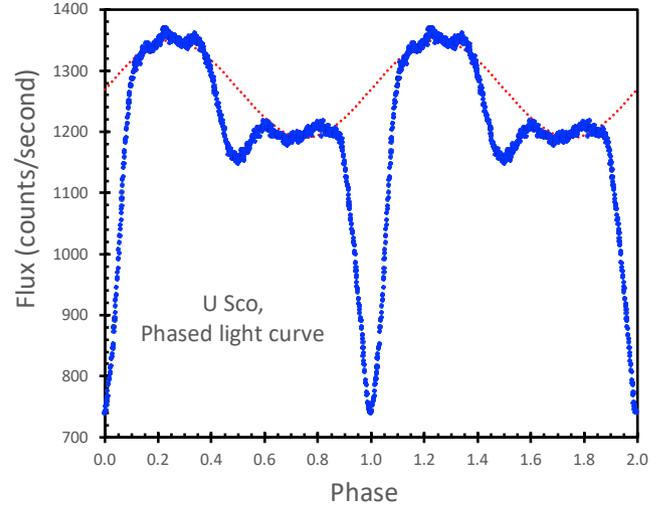}
    \caption{Phased and averaged light curve.  This light curve has all the K2 fluxes, with the long-term variations normalized out, folded on the best period, then averaged together into bins of size 0.001 in phase.  Each of the 1000 points with phase from 0--1 (repeated with phase 1--2) is composed of 107 fluxes averaged together, with each point having a one-sigma uncertainty close to 6.6 counts/second (i.e., smaller than the plot points).  All of the visible features and wiggles in the curve are significant.  The most prominent feature is the deep primary eclipse at phase 0.000 (and repeated at phase 1.000), as the G-type subgiant companion star eclipses most of the accretion disc.  The primary minimum has an asymmetry, where the ingress is relatively slow when compared to the egress.  Also visible is a secondary eclipse centred at phase 0.500 (and phase 1.500).  A prominent asymmetry appears in the out-of-eclipse light curve, where the phase from roughly 0.1--0.4 is $\sim$12 per cent brighter the star at the other quadrature (for phases 0.6--0.9).  The dotted red line is a sinusoid, as a suggestive description of the out-of-eclipse variations.}
\end{figure}

\begin{figure}
	\includegraphics[width=1.07\columnwidth]{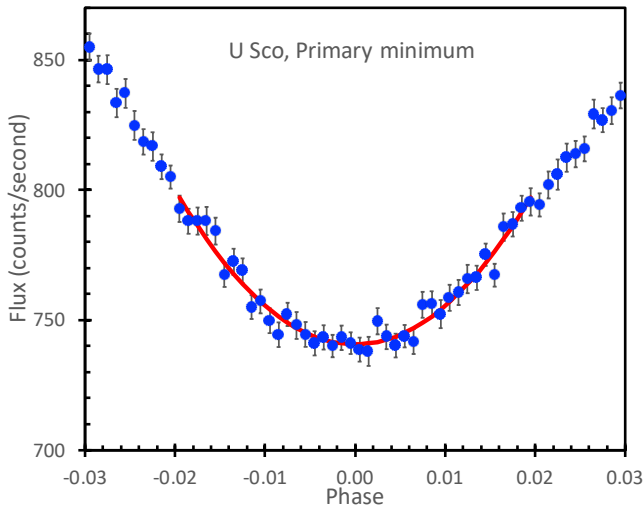}
    \caption{K2 light curve for the deepest primary minimum.  This light curve has a resolution of 0.001 in phase, with each datum being the average of the normalized fluxes from all the minima.  This is a blow-up of Fig. 4.  The light curve displays asymmetric differences that increase away from zero phase.  The light curve  looks to never have a perfectly flat segment at the deepest eclipse.  The best-fitting parabola (shown as a red curve) shows that the light curve is consistent with a parabolic minimum. }
\end{figure}

\begin{table}
	\centering
	\caption{K2 phased and averaged light curve for U Sco with 0.001 resolution, full table of 1000 fluxes is in the on-line supplementary material}
	\begin{tabular}{ll} 
		\hline
		Phase &  Flux (ct/sec)  \\
		\hline
0.0005	&	739	$\pm$	5	\\
0.0015	&	738	$\pm$	6	\\
0.0025	&	750	$\pm$	5	\\
0.0035	&	744	$\pm$	5	\\
0.0045	&	740	$\pm$	5	\\
...	&				\\
0.9955	&	741	$\pm$	5	\\
0.9965	&	743	$\pm$	5	\\
0.9975	&	740	$\pm$	4	\\
0.9985	&	743	$\pm$	5	\\
0.9995	&	741	$\pm$	4	\\
		\hline
	\end{tabular}
\end{table}

This light curve shows many features, all with exquisite detail and accuracy.  It appears to me that the basic shape is a sine wave (peaking around phase 0.25) with the primary and secondary eclipses superposed (see Fig. 4).  The sine wave would arise from the asymmetric emission of the disc light.  I speculate that this pattern arises from the hotspot, where the accretion stream impacts the accretion disc, making for a raised edge of the disc.  Around phase 0.25, the extra light from the irradiated inner side of the hotspot gas makes for a bright out-of-eclipse flux, while around phase 0.75 this same raised edge of the accretion disc is shadowing more of the inner accretion disc than at other phases.  A model light curve fit should determine the position of the hotspot (apparently $\sim$90$\degr$ around the disc circumference from the inner Lagrange point) and dynamics calculations for the accretion stream should then return the disc radius.  The disc radius should also be derived from the sharp edges of the primary and secondary eclipses (which appear to be close to 0.10 in phase away from the conjunctions of both primary and secondary eclipses for both ingress and egress).  A detailed light curve model should be able to key off the eclipse depths and durations to get the separation, inclination, and sizes of the disc and companion.  The U Sco phased light curve is by far the best such phased light curve for any nova that I am aware of, and that is why I have put the full thousand-point light curve into a table, as a challenge for modelers.

\subsection{Aperiodic variations}

U Sco is varying chaotically on all time-scales, with the periodic orbital variations superposed.  The K2 light curve shows rises and falls on time-scales of days, with the ordinary orbital modulations superposed.  A simple light curve has these long-term variations largely hidden by the orbital variations.  A means to hide the orbital effects is to construct a light curve where all the points are from the same orbital phase.  This can be quantified by a plot for each phase with one point for a given phase for each orbit.  Such a plot is shown in Fig. 6, for orbital phases of 0.00 (mid-eclipse), 0.25 (the bright quadrature), 0.50 (the middle of the secondary eclipse), and 0.75 (the faint quadrature)..  The red diamonds show the average flux over phases $-$0.005 to $+$0.005 for each  of 62 orbits.  The blue triangles record the flux averaged from phase 0.23 to 0.27, the bright quadrature after the primary eclipse.  The magenta squares record the phase range of 0.04 centred on the secondary eclipse at phase 0.50.  The green circles record the average flux from 0.73--0.77 for each orbit.  The formal error bars for each point are $\pm$6.5 counts/second for the 0.00 phase, and $\pm$3.3 counts/second for the other phases.  That is, the measurement and Poisson errors are greatly smaller than the plot symbol size in Fig. 6.  

\begin{figure}
	\includegraphics[width=1.1\columnwidth]{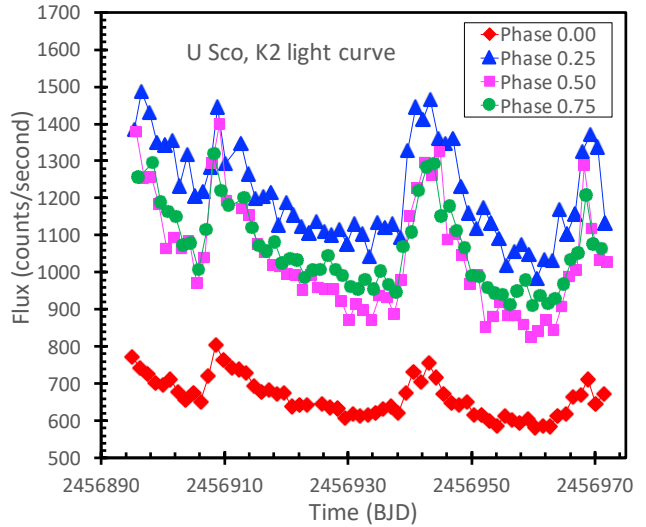}
    \caption{K2 light curve for four orbital phases. The point of this plot is to show the long-term variation in the brightness over 78 days. At least three flares lasting several days are visible.  These are apparently just part of a continuum of variability time-scales, as seen in many CVs.}
\end{figure}

The variations in the four phases go up-and-down in lockstep with each other.  It looks like the four curves are just scaled versions of each other, where the scale factor is from the beaming pattern of the disc light.  (That is, disc light will be beaming brighter in some directions, while eclipse effects can hide disc light in other directions.)  In this case, the fractional amplitudes should be the same for all four curves.  The companion star contributes substantial flux added to disc light, with this light being roughly constant.  Such extra light will reduce the fractional amplitude of the intrinsic variations of the disc light.  The reduction in amplitude will be greater for the 0.00 phase light curve than for the quadrature light curves. The observed amplitudes in Fig. 6 can be used to estimate the flux from the companion star.  The 0.00 phase light curve varies over a range 582--805 counts/second (28 per cent amplitude), while the 0.25 phase light curve varies over a range 984--1445 counts/second (32 per cent amplitude).  If the companion star contributes a constant flux of 205 counts/second, then the amplitudes of the disc component will be the same.  This estimate of the companion contribution will have substantial uncertainties because there are a variety of reasonable ways to extract amplitudes from the phased light curves, there are various combinations of phases that can be compared, and this simple model does not allow for variations in the companion light (from ellipsoidal effects, eclipsing of the companion, and irradiation effects).  The companion contribution of 205 counts/second can only be considered as approximate, where I cannot even calculate the size of the real uncertainty.  Still, this produces a reasonable quantitative measure of the contribution of the companion star light, in the K2 pass band.  Thus, the companion light is 14 per cent of the total system light when U Sco is at its brightest, and it contributes 35 per cent of the total system light when U Sco is at its faintest at mid-eclipse.

\subsection{Are the eclipses total?}

Are the primary eclipses total?  That is, does the companion star completely cover the white dwarf and accretion disc at the time of conjunction?  This question has importance for constraining the inclination, and for modeling the phased light curve.  If the eclipses are total, then photometry at the time of deepest eclipse can be used for measuring a blackbody distance to the companion star without uncertain corrections for disc light, and spectroscopy of the companion alone, with no illumination effects, will allow for standard analyses to measure the surface temperature, gravity, and composition.

The companion star and its Roche lobe are 2.3 R$_{\odot}$ in radius.  The estimated stellar masses for the white dwarf  and companion are 1.36 and 1.0 M$_{\odot}$, so the white dwarf Roche lobe has a radius of around 4.2 R$_{\odot}$.  The radius of the accretion disc must be smaller than the white dwarf's Roche lobe, with the exact ratio of the radii being poorly known.  The minimum disc radius is the circularization radius (c.f., Frank et al. 2002, eq. 4.20), which is around 0.9 R$_{\odot}$ for U Sco, with the real disc perhaps being substantially larger.  In all, we can only constrain the radius of the disc to be somewhere near the middle of the possible range of 0.9--4.2 R$_{\odot}$.  This is comparable to the companion star radius.  So even for an edge-on system, we cannot predict whether the U Sco eclipse should be total or partial.

Here, I marshal three strong arguments that demonstrate that the eclipse is {\it not} total, which is to say that the disc peeps out on both sides of the companion star at conjunction:

{\bf (1)~}The hallmark of a total eclipse is a flat-bottomed light curve.  Schaefer (2011) has already shown that the eclipse shape is close to flat over the phase range from $-$0.010 to $+$0.010.  Within this small range, the K2 light curve has 2200 fluxes, each with typically 3.6 per cent measurement errors, and this makes a substantially improved measure of the shape of the bottom of the eclipse.  The normalized and folded K2 light curve with 0.001 phase bins is shown with a close-up around the mid-eclipse in Fig. 5. The minimum is fairly flat between phases $-$0.01 and $+$0.01, but that some curvature is still visible.  I have formalized this by-eye result with a chi-square fits for models of parabolas with cutoff flat bottoms.  These fits have the lowest chi-square for the case with no cutoff.  When the cutoff is raised to a level with a flat bottom between phases $-$0.010 to $+$0.010 the chi-square rises above the minimum value to a degree corresponding to a 4-sigma rejection.  Thus, U Sco primary minimum does not have a flat bottom, and the eclipse is not total.


{\bf (2)~}If the eclipse is total, then the flickering must vanish throughout totality.  This can be tested by looking at the variations of the K2 light curve (Fig. 3) around the phase-averaged folded light curve (Fig. 4).  Flickering flares will reveal themselves as short term deviations above the average folded light curve.  These deviation show flickering at zero phase, with no apparent eclipsing.  A plot the RMS scatter of the light curve deviations in small phase bins shows the statistical properties of the flickering as a function of phase.  The average RMS outside-of-eclipse is 67 counts/second, while the average RMS within 0.01 of zero phase is 49 counts/second.  After subtracting out the variance from the Poisson noise, the RMS outside eclipse is 61 counts/second, while the RMS at mid-eclipse is 41 counts/second.  This demonstrates that flickering is present at mid-eclipse, which proves that the eclipse is not total.  Further, roughly half the variance due to flickering alone must come from the outer regions of the disc, and half the variance from the inner regions, proving that flickering simultaneously arises in both the inner and outer disc..

{\bf (3)~}If the eclipse is total, then at mid-eclipse only the un-illuminated backside of the companion is visible, and the system should always have the same brightness, regardless of the highly variable accretion rate.  For the homogenous set of 22 eclipse light curves from G. Myers (MGW), the parabola-fitted minimum $CV$ magnitudes vary from 19.71 to 18.92 mag.  The central 50 per cent of this distribution has a range of 19.40--19.24 mag.  For the homogenous set of 59 eclipses observed with K2, the bottom curve in Fig. 2 shows the variation of mid-eclipse brightnesses varying  from 582 to 805 count/second.  Here is proof that disc light is prominent at conjunction, which is to say that the eclipse is not total.


\subsection{Eclipse times}

The K2 mission has produced beautiful light curves with good time resolution and good photometric accuracy, covering 59 eclipses in late 2014.  Each of these light curves had the minimum fitted to a parabola, covering the faintest half-magnitude or so.  The critical product is the time of minimum light, along with its one-sigma uncertainty.  Details are the same as for the ground-based eclipse times (Section 2.2) and the pre-2011 eclipse times (Schaefer 2011).   The eclipse times are presented in Table 4.

These K2 eclipse times represent a very-homogenous and accurate sample, all for 59 eclipses all in a row.  As a long-time observer and analyst of the U Sco eclipses, I was initially surprised by the wide scatter of the $O-C$ for the K2 times.  With the wonderful stability, accuracy, and homogeneity over just 78 days, I was expecting the $O-C$ to be a constant with little scatter.  Instead, apparently random jitter up-and-down is at a level far above the measurement errors.  Specifically, the RMS scatter of the $O-C$ is 0.0036 days, whereas the average one-sigma measurement error is 0.0011 days.  The jitter is being dominated by some other source of scatter, with this having an RMS of $\sqrt{(0.0036^2-0.0011^2)}$ or 0.0034 days.  This dominating source of jitter in the timings cannot realistically be attributed to unrecognized measurement error, nor to the results of any inhomogeneity in the data, nor to any instability in the instrument or analysis, nor to small number statistics.  All the usual possibilities are certainly eliminated in this very clean data set.

Schaefer (2021) came up with a previously unrealized mechanism to explain the jitter in the U Sco eclipse times, plus similar jitter in other CVs.  The jitter is a simple consequence of ordinary flickering.  That is, a small amplitude flicker on the ingress will shift the fitted parabola to a time somewhat after conjunction, while a small amplitude flicker on the egress will shift the fitted parabola to a time somewhat before conjunction.  The U Sco flickering appears throughout the depths of the eclipse (Section 3.3), and this mechanism is inevitable and ubiquitous.  I have simulated this mechanism for the case of U Sco.  In each simulation, I have taken the average eclipse light curve (see Fig. 5), added on a typical flicker, used my standard parabola fit to determine the time of minima, and then compiled the deviations from the zero phase of the average light curve.  The deviations of course depend critically on the timing, amplitude, and duration of the added flicker.  For typical U Sco flickers, the deviations in time-of-minimum vary back-and-forth with a range comparable to the 0.0034 day jitter.  This demonstrates that the timing jitter from ordinary flickers can easily account for the observed scatter in the $O-C$ as seen by K2.  So I conclude that the observed $O-C$ jitter is an expected consequence of ordinary flickers shifting the time-of-minimum back-and-forth with respect to the time-of-conjunction.

This result shows that the measurement accuracy of eclipse times in CVs has an irreducible random jitter that is intrinsic to the star.  That is, no matter the size of the telescope and no matter the quality of the light curve, each and every eclipse time will randomly jitter around the true time of conjunction by some constant.  This jitter is intrinsic to the star, and the aperiodic random nature of the flickering means that there is no way of correcting for the individual flickers.  All observers are stuck with this irreducible error.  For U Sco, the jitter from flickering has an RMS of 0.0034 days.  

This realization has many implications for observers and theorists:  {\bf (1)} One practical application for theorists is that they should stop fitting the jitter noise in $O-C$ curves.  That is, previously, theorists would fit a CV $O-C$ curve dominated by flickering jitter, variously to parabolas and sine waves, then make deductions about the physics of the binary.  {\bf (2)} A second practical application is for understanding the real total uncertainty, where the  total uncertainty in each eclipse time is from the addition in quadrature of the one-sigma measurement error bar (as reported in Table 4) plus the flicker jitter of 0.0034 days.  For chi-square fits to the $O-C$ curve, the individual times should be used, along with their realistic error bars, with these producing best-fittings in the $O-C$ curve with reduced chi-squares near unity.  {\bf (3)} Another practical application is the realization that the only way to beat down the errors is to measure many eclipse times, each with its own small random shift from flickering.  The average $O-C$ for the 59 K2 eclipse times will produce a combined value with an uncertainty of around 0.0034 days divided by $\sqrt{59}$, or 0.0004 days.  The observational imperative is to observe as many eclipses as possible with most any telescope.  {\bf (4)} A fourth practical application is that the $O-C$ curve is best displayed with seasonal-averaged values, to avoid the vertical cloud of plotted values that hides the good accuracy of the average.  {\bf (5)} Another practical application is that little is gained, on this issue, by having very-high-quality light curves. The final total uncertainty from `small' amateur telescopes is just as good as from the vaunted {\it Kepler} spacecraft, for the applications in this paper.  {\bf (6)} A sixth practical application is that having many eclipse times from one observing-season is good, but for questions about period changes in a CV, the need is for many seasons.  For all the key questions for U Sco concerning period-changes (see the next Section), what matters most is getting coverage over many seasons.  With this, the many seasons with good eclipse times by Myers is greatly more useful than all the 78 days of continuous coverage by {\it Kepler}.

\section{$O-C$ CURVE}

The $O-C$ curve is the long traditional plot for showing the orbital period and its changes, both for observed eclipse times and for theoretical models.  ``$O-C$'' stands for `observed minus calculated' for some phase marker.  For CVs, the adopted zero phase is the time of conjunction when the white dwarf is farthest from Earth, which in cases like U Sco is taken to be the time of minimum light in the primary eclipse.  The observed time can be notated as $T_N$, where the subscript $N$ identifies the particular eclipse by a running integer counting the orbits from some fiducial epoch.  For U Sco, the fiducial epoch is an eclipse time at the start of the 1999 eruption, taken as HJD 2451234.5387.  The period is well enough known such that the cycle count can be confidently known for many decades before and after 1999.  The calculated time (the `$C$' in $O-C$) is determined from some ephemeris, almost always a linear ephemeris with a constant period.  For U Sco, I will adopt the linear ephemeris used in earlier publications (e.g., Schaefer 2011).  The eclipse times predicted by the fiducial ephemeris are given as
\begin{equation}
T_{eph} = 2451234.5387 + N\times 1.23054695.
\end{equation}
The observed eclipse times, $T_N$, will differ from this schematic model, and the difference is the $O-C$ value, 
\begin{equation}
O-C = T_N - T_{eph}.
\end{equation}
A plot of $O-C$ values versus time is the $O-C$ curve.  At any time, the slope of the $O-C$ curve gives the current period $P$ relative to the fiducial period in equation (4).  The $O-C$ curve must be continuous, with no vertical discontinuities, which is just to say that the companion star does not hop around in its orbit.  The {\it slope} can have discontinuities, for example, from sudden period changes at the time of the nova event making a sudden loss of mass in the system.  The slope can also have slow and steady changes throughout the inter-eruption quiescence, resulting in a parabola shape in the $O-C$ curve.  Such linear period changes can be caused by a number of mechanisms, including from magnetic braking of the companion star's rotation or from ordinary mass transfer.  It is exactly all these period changes that are driving CV evolution, thus making for the critical importance  of measuring the long-term $O-C$ curves of the few novae for which it is possible.  

An $O-C$ curve can also be constructed for some theoretical or fitted model, $T_{model}$, with 
\begin{equation}
O-C = T_{model} - T_{eph}.
\end{equation}
For example, the model might predict that the period changes like a sine wave (say, from a third body in orbit around the binary), or the model might invoke a linear period change (say, from magnetic braking), or the model might invoke a sudden period change (say, from mass loss in a nova eruption).  These three period behaviours will produce an $O-C$ curve that displays a sine wave, a parabola shape, and a sharp kink, respectively. The U Sco $O-C$ curve will be composed of parabola segments and kinks at the time of eruptions.  Indeed, the primary science of this paper is to measure any curvature and kinks in the U Sco $O-C$ curve.

Between eruptions, period changes at some steady rate result in a parabolic segment in the $O-C$ curve.  As an equation, this can be expressed as 
\begin{equation}
T_{model} = E + NP  + 0.5  N^2 P \dot{P}.
\end{equation}
Here, $E$ is the epoch of some particular eclipse time, which I will always take to be at the start of a U Sco eruption.  $N$ is the running count of orbits since the epoch, while $P$ is the period immediately after the eruption.  The steady period change is $\dot{P}$, which in this equation is dimensionless.  This formulation for a dimensionless $\dot{P}$ is to match the usual expression of theory models.

The U Sco situation is unique in having sudden period changes measured across {\it three} eruptions (in 1999.2, 2010.1, and 2016.78), plus different $\dot{P}$ values in each of four inter-eruption interval.  I will use the basic parabola in equation (7) for each of the four intervals between eruptions.  Each interval will be designated with a letter subscript, with the interval from 1987.4 to 1999.2 identified with `A`, and so on (see Table 7).  In the model, each eruption time will have a particular value corresponding to an eclipse time at the start of the eruption ($N_A$, $N_B$, and so on), and the corresponding epoch ($E_A$, $E_B$, and so on), as listed in Table 7.  The model epochs are free parameters for fitting to the data. Four parabola equations are needed to express the model over the four inter-eruption intervals:
\begin{multline}
T_{model} = \\  ~~~~~E_A + (N-N_A)P_A  + 0.5  (N-N_A)^2 P_A \dot{P}_A,~~~ -3499<N<0  \\
~~~~~E_B + (N-N_B)P_B  + 0.5  (N-N_B)^2 P_B \dot{P}_B,~~~ 0<N<3243  \\
~~~~~E_C + (N-N_C)P_C  + 0.5  (N-N_C)^2  P_C \dot{P}_C, ~~~3243<N<5233  \\
~~~~~E_D + (N-N_D)P_D  + 0.5  (N-N_D)^2 P_D \dot{P}_D,~~~ 5233<N  \\
\end{multline}
The cycle count $N$ acts like a unit of time, although $O-C$ curves are most usefully plotted with time in units of the year.  This model for the 1989--2022 eclipse times has 12 fit parameters.  Three of the fit parameters can be eliminated by the requirement of continuity across eruptions.  With this equation and the best-fitting parameters, period can be calculated for just before an eruption, and just after the eruption.  Quantities of interest are the total period change throughout each interval of quiescence ($\Delta P_{q}$), the sudden period change across each nova eruption ($\Delta P_{nova}$), and the total period change over an entire eruption cycle ($\Delta P_{cycle}$=$\Delta P_{q}$+$\Delta P_{nova}$).

I have made a chi-square minimization for the 151 eclipse times outside of eruption (Table 4), covering 1989 to 2022.  The best-fitting parameters are placed into Table 7.  This fit has a chi-square of 140.49 for 142 degrees of freedom.

\begin{table*}
	\centering
	\caption{Period change results from $O-C$ diagram fits}
	\begin{tabular}{lllll}
		\hline
		  & A: 1987.4--1999.2 & B: 1999.2--2010.1  & C: 2010.1--2016.78   &   D: 2016.78 to 2022.4 \\
		\hline
$Y_{start}$ (year)	&	1987.4			&	1999.2			&	2010.1			&	2016.78			\\
$N$	&	$-$3499			&	0			&	3243			&	5233			\\
$E$ (HJD)	&	2446928.8310	~$\pm$~	0.0022	&	2451234.5346	~$\pm$~	0.0032	&	2455225.2026	~$\pm$~	0.0013	&	2457673.9930	~$\pm$~	0.0021	\\
$P$ at $Y_{start}$ (days)	&	1.2305596	~$\pm$~	0.0000038	&	1.2305504	~$\pm$~	0.0000031	&	1.2305737	~$\pm$~	0.0000025	&	1.2305658	~$\pm$~	0.0000041	\\
$\dot{P}$ ($10^{-9}$)	&	$-$3.2	~$\pm$~	1.9	&	$-$1.1	~$\pm$~	1.1	&	$-$21.1	~$\pm$~	3.2	&	$-$8.8	~$\pm$~	2.9	\\
$\Delta P_q$ ($10^{-6}$ days)	&	$-$14.0	~$\pm$~	8.2	&	$-$4.3	~$\pm$~	3.9	&	$-$51.5	~$\pm$~	8.0	&	$-$18.1	~$\pm$~	6.0			\\
$\Delta P_{nova}$$(10^{-6}$  days)	&	$+$4.8	~$\pm$~	7.6	&	$+$27.6	~$\pm$~	1.2	&	$+$43.6	~$\pm$~	8.7	&	...			\\
$\Delta P_{cycle}$ ($10^{-6}$ days)	&	$-$9.2	~$\pm$~	11.1	&	$+$23.3	~$\pm$~	4.1	&	$-$7.9	~$\pm$~	11.8	&	...			\\
		\hline
	\end{tabular}
\end{table*}

\subsection{Sudden period change near 2016.78}


Fig. 7 shows the observed $O-C$ curve from 2009--2023, with the seasonal values plotted as blue diamonds.  The chi-square for the two inter-eruption intervals from 2010--2020 is 97.23 from the 103 individual eclipse times, all with 4 free fit parameters.  One of the prominent features, especially as I was watching this unfold in real time, is the `deep' parabola from 2010.1--2016.78.  A parabola is expected after the 2010.1 eruption, but the high curvature was startling because it is much more curved than prior times of quiescence (see Schaefer 2011).  This provides immediate proof that CV $\dot{P}$ values can quickly change by large factors, in this case across a nova event, with this being unprecedented, unanticipated, and still inexplicable.  No known mechanism can change $\dot{P}$ on any fast time-scale, hence the entire time between eruptions should have a single parabola for its $O-C$ curve.  The best-fitting parabola is displayed in Fig. 7, with an extension past 2016.78 as shown in the dashed red curve.

Fig. 7 has the important point that the observed $O-C$ curve shows a strong kink at near 2016.78.  That is, the times from 2019 and afterwards are far above any extended parabola.  For example, the well-measured parabola when extrapolated to 2019.4 is for an $O-C$ value of $-$0.0242 days, whereas 8 eclipses in that season have an average value of $+$0.0135$\pm$0.0014 days.  This difference is 54 minutes and 27-sigma, and the errors get more extreme in later years.  This is the proof of the missing nova event.

Fig. 7 also shows the 2016.78--2022 observations.  These show a steady progression with a modest amount of curvature, just as expected for an ordinary post-eruption interval.  Critically, the post-2016.78 $O-C$ curve points back to the well-defined 2010.1--2016.78 parabola for a time  around late 2016, which I fit to be 2016.9$\pm$0.6.

\begin{figure}
	\includegraphics[width=1.05\columnwidth]{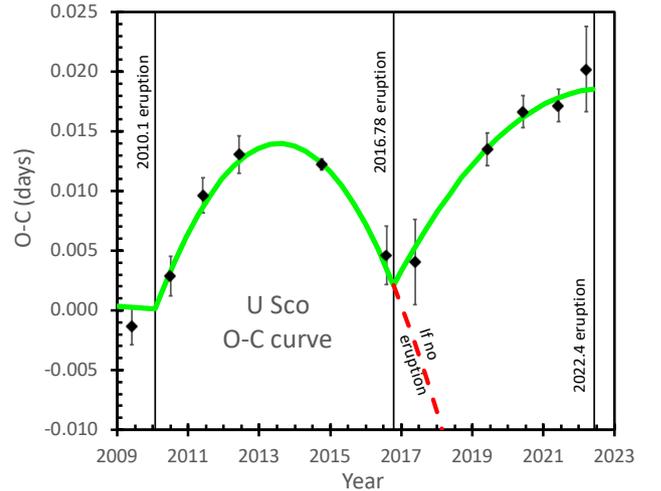}
    \caption{The $O-C$ curve has a highly significant kink near 2016.78.  After the 2010.1 eruption, the eclipse times closely followed a parabola (the green curve), at least up until 2016.78.  After 2016.78, in the absence of any nova eruption, the $O-C$ would continue down as an extension of the well-defined parabola, as depicted by the red dashed curve.  This possibility is contrary to the observations.  Rather, the $O-C$ has a sudden and sharp deviation from the extended parabola, such that a completely different parabolic segment continues after the 2016.78 kink.  This sharp kink is characteristic of a nova eruption, and I know of no other possibility for producing such a fast period change in the middle of a time interval of quiescence.   This is the proof that U Sco had a lost eruption late in the year 2016.}
\end{figure}

\subsection{Large changes in $\Delta P_{nova}$ and $\dot{P}$ from eruption-to-eruption}

Up until the time of the 2010.1 eruption, the U Sco $O-C$ curve only displayed moderate curvature and a $\Delta P_{nova}$ that was consistent with zero (Schaefer 2011).  Starting in 2011, I was startled that my eclipse timings from CTIO had such a high $O-C$.  I ran through a variety of systems checks, for example by testing the clocks used at CTIO.  Then with my 2012 eclipse times running even higher, there was no getting around the fact that U Sco suffered a very large positive $\Delta P_{nova}$ across the 2010.1 eruption.  When the K2 eclipse times came in, there could be no arguing from the fact that the post-2010.1 $O-C$ curve was showing a very large negative $\dot{P}$.  And this was confirmed with my 2016 eclipse times from CTIO.  From 2010.1--2016.78, U Sco had a near-perfect parabola in its $O-C$ curve, and the $\Delta P_{nova}$ and $\dot{P}$ were provably greatly larger than previous measures.  All of the $\Delta P$ and $\dot{P}$ measures from 151 out-of-eruption eclipse times have been collected in Table 7, with the best-fitting model for equation (8) displayed in Fig. 8 as a thick green curve.

For first two inter-eruption intervals, the $\dot{P}$ has substantial uncertainties.  For the interval before the 1999.2 eruption, the 17 eclipse times have relatively large scatter about any smooth O-C curve.  Part of this problem is that half the eclipses were truncated in time and miss parts of either the ingress or egress.  (This was due to not knowing the eclipse times in advance for the first five eclipses, and then in later years by being constrained by the available nights of telescope time not having optimally timed eclipses.)  The problem is also apparent with the one eclipse time in 1990 having a greatly different $O-C$ from the average for the five eclipse times in 1989.  The single 1990 eclipse time was derived from measures of the continuum flux of 8 spectra throughout the eclipse, in a procedure unique amongst all the eclipse times in Table 4.  It is easy, and perhaps tempting, for some outsider to arbitrarily reject either or both the 1989 or 1990 averaged $O-C$ values to achieve the smoothest $O-C$ curve.  But such a temptation is wrong, as the answer (to the value of the earliest $\dot{P}$) would be based on throwing out data with no cause.  Rather, the measures and their error bars are reasonable and unimpeached, hence all the data must be used as quoted, as with the chi-square fits.

Fig 8 and Table 7 show that both $\Delta P_{nova}$ and $\dot{P}$ are changing greatly from eruption-to-eruption.  The obvious question is to ask how the three eruptions (1999.2, 2010.1, and 2016.78) differ from each other, and the answer is that they are identical in light curve (Schaefer 2010; Pagnotta et al. 2015) and in spectral development.  The root cause of the eruption-to-eruption changes is somehow invisible.   The eruption-to-eruption period changes are startling because there has been no precedent, no worker has made prediction or any expectation that such period changes would occur, no published models even addresses the question, and I am not aware of any mechanism in a CV that could possibly make for such changes.

\begin{figure}
	\includegraphics[width=1.1\columnwidth]{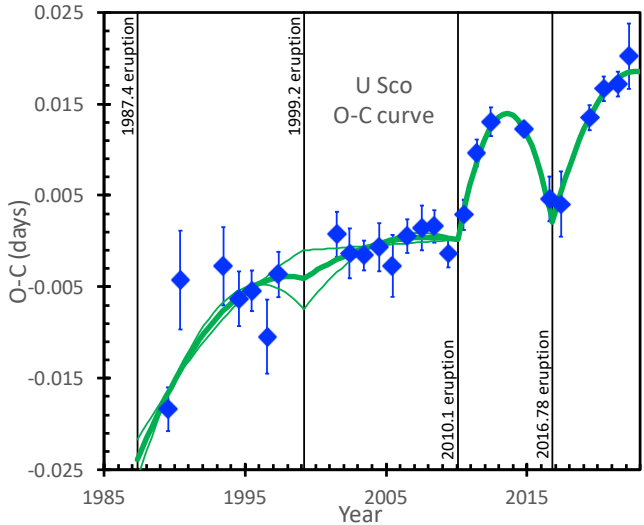}
    \caption{The $O-C$ curve from 1989 to 2022.3.  The $O-C$ curve consists of parabolic segments (caused by steady period changes during quiescence) that have sharp kinks at the times of nova events (caused by mechanisms like mass loss).  The thick green curve is the best-fitting $O-C$ curve, from equation (8) and with the parameters in Table 7.  This is flanked by curves that are one-sigma different in the chi-square, with the difference primarily being visible before the 2010.1 eruption.  These flanking curves show that the behaviour around the 1999.2 eruption is poorly constrained, the period change has a substantial uncertainty, and might be either positive or negative.  The main point of this figure is to see that the $\Delta P_{nova}$ and $\dot{P}$ values are changing from eruption-to-eruption.  That is, the period change across the 1999.2 eruption is small and consistent with zero, while both the 2010.1 and 2016.78 eruptions have very large and positive period changes.  This is the first measure of $\Delta P$ across more than one eruption of the same star, and there is no precedent or expectation for such large eruption-to-eruption differences in $\Delta P$.  The steady period changes between eruptions is also changing greatly from eruption-to-eruption.  This is apparent by the fairly flat segment from 1999.2--2010.0 versus the strong curvature from 2010.1--2016.78.  There is no precedent, expectation, or understanding for the eruption-to-eruption differences in $\dot{P}$.}
\end{figure}

\subsection{Long-term evolution from 1906--2022}

I have accurate measures of $\Delta P_{nova}$ for three eruptions and $\dot{P}$ across four inter-eruption intervals.  Plus, I have eclipse times from 1945 to 2022.  Further, as measures of the accretion rate, I have $\langle V \rangle$ values from 1976 to 2022.  Plus, with deep and wide searching of archival data, I have lists of eruption dates (giving average accretion rates) that is likely complete going back to 1906.   These long ranges of measures have averaged over the various eruption-to-eruption changes, and the long-term evolution effects should be visible.

The evolution of CVs is driven by the period changes.  Old CV evolution models have idealized the evolution as being driven by the $\Delta P_{nova}$ (for the Hibernation model) or being driven by the steady $\dot{P}$ (for the Magnetic Braking model).  Now, with U Sco, {\it both} are of comparable importance.  In particular, despite substantial eruption-to-eruption variability, the average values of $\Delta P_{nova}$ and $\dot{P}$ are approximately offsetting.  To pull out the effects for three complete cycles, the interval from 1994 to 2022 is well measured.  (This interval is from mid-cycle to mid-cycle, chosen to avoid the worrisome-but-unimpeached eclipse times from 1989 and 1990.)  The three eruptions give a total of $\Delta P_{nova}$ = ($+$76$\pm$12) $\times 10^{-6}$ days.  The interval from 1994 to 2022 has a total of $\Delta P_{q}$ = ($-$78$\pm$12) $\times 10^{-6}$ days.  That is, the sudden period increases across nova events is cancelled out by the steady period decreases throughout the quiescent intervals.  For this three-cycle interval, $\Delta P_{cycle}$ = ($-$2$\pm$17) $\times 10^{-6}$ days.  Alternatively, for the three-cycle interval from just after the 1987.4 eruption to just after the 2016.78 eruption, the values in Table 8 give $\Delta P_{nova}$ = ($+$76$\pm$12) $\times 10^{-6}$, $\Delta P_{q}$ = ($-$70$\pm$12) $\times 10^{-6}$, and $\Delta P_{cycle}$ = ($+$6$\pm$17) $\times 10^{-6}$ days.  The point is that the average $\Delta P_{nova}$ and $\dot{P}$ values are quite large in size, but they come near to canceling each other out.

We can do better, over a much longer time interval, by using the 1945 eclipse from the Harvard plates (Schaefer 2011).  The 1945 eclipse time was extracted from 19 plates taken late in the tail of the 1945 eruption, for which 3 plates on 1945 July 2 show U Sco in eclipse, being 1.0 mag below the detrended out-of-eclipse level (see fig. 9 of Schaefer 2011).  There is no doubt about the existence of the eclipse, although the eclipse time has the relatively large uncertainty of $\pm$0.0090 days.  The 1945 eclipse is 34 days after the eruption peak.  Based on the many well-measured eclipse times throughout the tail of the 2010.1 eruption, the $O-C$ at a time 34 days after peak shows a near-zero offset to the nearby times in quiescence, hence the reported eclipse time can be treated as if it were a time in quiescence.  

A concern is the cycle count from 1945 to 1989.  With the fiducial ephemeris in equation (4), the cycle count back to the 1945 eclipse gives $N$=$-$15924, and then its $O-C$ is very close to zero ($-$0.0091$\pm$0.0090 days).  If the cycle count were really off by one, then the $O-C$ would be either $-$1.2396 of $+$1.2214 days.   This can be evaluated in three ways:  {\bf (1)} If the cycle count is wrong, then it is rather improbable, at the 0.7 per cent probability level, that equation (4) would produce an $O-C$ value close to zero, and this is a good argument that the $-$15924 count is correct.  {\bf (2)} For the cycle count of $-$15924, a fitted parabola, representing a smooth evolution, from 1945--2022 (see Fig 9) reasonably describes the measured $O-C$ measures, even with the fast period changes.  But if we assume the cycle count is $-$15923, then to pass through the 1945 eclipse time (with $O-C$=$-$1.2396 days), the required $\dot{P}$ is large and negative ($-$5.9$\times$10$^{-9}$) applied to the entire time interval, with this making for a chi-square that is 25$\times$ larger than in the $N$=$-$15924 case.  To get the $O-C$ curve to pass through the 1945 measure, the best fit parabola goes from below the Fig. 9 plot range in 2022, to far above the plot range from 2013 and back to 2003, to below the plot range in 1995, and continuing down going back to 1945.  That these contortions disagree so strongly with the 1989--2022 $O-C$ curve demonstrates that the cycle count of $-$15923 is not reasonable.  A similar analysis rejects the cycle count of $-$15925.  {\bf (3)} Alternatively, we could ask what average $\Delta P_{nova}$ is needed for the 1957, 1969, 1979, and 1987 eruptions so that the $O-C$ passes through $-$1.2396 back in 1945.  To make the cycle count of -15923 back in 1945, all four eruptions would have to have an average $\Delta P_{nova}$ equal to $-$32 in units of $10^{-6}$ days.  This possibility seems poor when we recall that the three actual observed values are $+$4.8, $+$27.6, and $+$43.6, all with the opposite sign.  So we can be confident that the cycle count to the 1945 eclipse is the obvious $-$15924.

\begin{figure}
	\includegraphics[width=1.1\columnwidth]{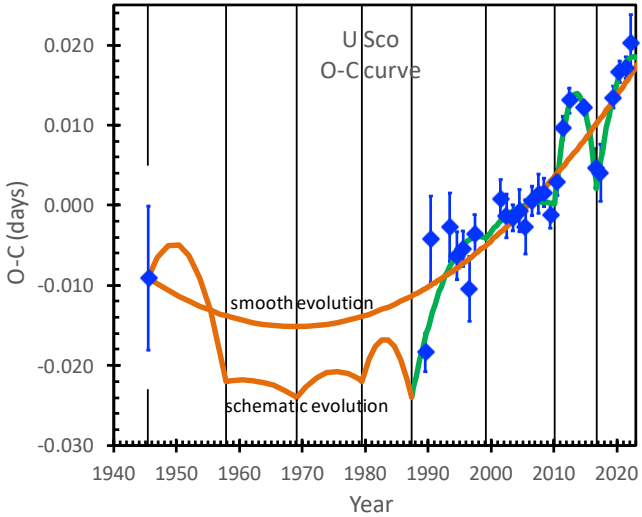}
    \caption{The $O-C$ curve from 1945 to 2022.  This $O-C$ curve adds on the critical 1945 eclipse from the Harvard plates.  Now, the 77 years (covering 7  observed eruptions) averages over the variations to get a view of the long-term evolution in $P$ for U Sco.  The period evolution from 1945 to 1987 will consist of parabola segments, with the breaks at the times of nova events (indicated by the thin vertical black lines).  As an illustration, a possible schematic curve is shown as a serrated orange curve.  The long-term evolution, with smoothing over each eruption cycle can be represented with a single parabola from 1945--2022 (shown as the orange parabola labelled as ``smooth evolution").  This best-fitting parabola has a $\dot{P}$ of $+$0.20$\pm$0.05 $\times$ 10$^{-9}$, with this being the best number to describe the long-term evolution of the U Sco $P$.}
\end{figure}

The $O-C$ curve from 1945 to 1989 consists of parabola segments pieced together.  The serrated orange curve in Fig. 9 shows a schematic evolution, just using typical period changes.  This is presented mostly to impress on the reader the situation for the period changes.  The extension back to 1945 cannot tell us about $\Delta P_{nova}$ or $\Delta P_{q}$ individually, but it can tell us $\Delta P_{cycle}$ over the entire 1945--2022 interval.  In particular, the 1945 eclipse time forces the general shape of the $O-C$ curve to curve upwards when going back in time from 1989.  This is illustrated in Fig. 9 by the fitted single parabola, shown in orange, from 1945 to 2022, where this models the evolutionary results averaged over the eruption cycles.  With the 1945 eclipse, the long-term evolution of U Sco has a secular {\it increase} of the orbital period.  The parabola has a $\dot{P}$ equal to $+$0.20$\pm$0.05 in units of 10$^{-9}$.  This is the best number for trying to understand the long-term evolution of $P$ in U Sco.

We can also consider the long-term evolution of the accretion rate.  This can be measured by the $\langle V \rangle$, as in Fig. 2 and Table 5.  The chaotic variations are up and down on time-scales comparable to the inter-eruption cycle, but with no  apparent secular trends.  That is, the accretion rate was relatively very high from 2010.1 to 2016.78 (explaining the very short $\delta Y$), the accretion rate was relatively high from 1979.5 to 1987.4 as based on sparse data (explaining the short $\delta Y$=7.9 years), and the accretion rate is below average after the 2016.78 eruption (hence a prediction for a long $\delta Y$).  While $\langle V \rangle$ and hence $\dot{M}$ certainly controls $\delta Y$, I see no apparent correlation with changes in $\Delta P_{nova}$ or $\dot{P}$.

We have a longer record of the accretion rate averaged between eruption from the observed inter-eruption interval, with $\langle \dot{M} \rangle$~$\propto$~$\delta Y$$^{-1}$.  Known U Sco eruptions are in 1906, 1917, 1936, 1945, 1969, 1979, 1987, 1999, 2010, and now 2016.78 plus 2022.4.  Undoubtedly, the expected lost eruptions were around 1927 and 1957 (Schaefer 2010).  This leaves $\delta Y$ values of 10.8, $\approx$19.3/2, $\approx$19.3/2, 8.9, $\approx$23.7/2, $\approx$23.7/2, 10.4, 7.9, 11.8, 10.9, 6.68, and 5.6 years in order.  Inverting these values to get a quantity that scales as the average accretion rate, gives 0.093, $\approx$0.104, $\approx$0.104, 0.112, $\approx$0.084, $\approx$0.084, 0.096, 0.127, 0.085, 0.091, 0.149, and 0.18.  I see no significant secular trend.  The $\langle \dot{M} \rangle$ is not evolving significantly on time-scales up to 110 years.

\section{CRITICAL IMPLICATIONS OF THE PERIOD CHANGES}


\subsection{The missed eruption of 2016.78}

The simple $\sim$10 year recurrence time for U Sco, already led to an expectation back in 2010 for an upcoming nova event around the year 2020, with an uncertainty of a few years.  In 2018, the analysis of the quiescent brightness showed an anomalously bright U Sco after the end of the eruption in early 2010, with this demonstrating that the eruption had already past.  A detailed analysis shows that the U Sco white dwarf accreted $M_{trigger}$ in 2017.1$\pm$0.6 (see Section 2.5).  An analysis of the quiescent light curve shows a solar gap that allows an eruption peak only during a 75-day interval centred on 2016.78 (Section 2.6).  Then in 2019,  eclipse times revealing a sharp kink close to 2016.9, and this could only have come from a lost-eruption.  Subsequent eclipse times and analysis, in this paper, demonstrate a kink in the $O-C$ curve in 2016.9$\pm$0.6.  Now, the three independent means set the date of the lost eruption to be 2016.78$\pm$0.10.  In the end, the U Sco period changes are what provided the proof of the lost eruption around 2016.78.

\subsection{Variable $\Delta$P is apparently caused by asymmetric ejecta}

The primary result of my 1989--2022 program has been the measure of three $\Delta P_{nova}$ values for U Sco.  The $\Delta P_{nova}$/$P$ values are $+$3.9$\pm$6.1, $+$22.4$\pm$1.0, and $+$35.4$\pm$7.1 parts-per-million (ppm) for the three measured nova events.  These values are startling, because they are greatly smaller than prior expectations from models (the Hibernation model in particular), because they are greatly larger than possible by any of the well-known mechanisms, and because the $\Delta P_{nova}$ values change by an order-of-magnitude from eruption-to-eruption.  Here, let me consider the three well-known mechanisms for sudden period changes across a nova, plus one more mechanism only sketched briefly in the past literature:

{\bf (1)~~}The first and best-known mechanism to change the orbital period is the simple and inevitable fact that the mass loss from the white dwarf must also produce a substantial change in the orbital period.  That is, from Kepler's Law, as the white dwarf mass, $M_{WD}$, suffers a sudden drop for the ejected mass, the $P$ must also change.  The fractional period change due to mass loss is close to 
\begin{equation}
\Delta P_{ml}/P = 2 M_{ejecta} / (M_{WD}+M_{comp}).
\end{equation}
We can adopt the value $M_{WD}$=1.36 M$_{\odot}$ (Shara et al. 2018), and the approximate companion star mass is $M_{comp}$=1.0 M$_{\odot}$.  The nova ejecta mass is very poorly known, with observational estimates ranging from 0.01 to 7.2 in units of $10^{-6}$ M$_{\odot}$, while theory estimates range from 0.21 to 4.4 in the same units (Appendix A in Schaefer 2011), with some sort of a middle estimate being $M_{ejecta}$$\sim$10$^{-6}$ M$_{\odot}$.  The ejecta mass cannot be larger than 2.5$\times$10$^{-6}$ M$_{\odot}$, as the white dwarf would have to be receiving mass for 10 years at a rate that stable hydrogen burning on the surface would prevent nova eruptions (Shen \& Bildsten 2009).  For the middle value of the ejecta mass, $\Delta P_{ml}$/$P$ is 0.8 ppm.  Even for the maximum ejecta mass, $\Delta P_{ml}$/$P$ equals 2.1 ppm.  With this, the observed $\Delta P_{nova}$/$P$ value for the 1999.2 eruption can only be explained only by pushing the theory value to its maximum while pushing the observed value to the lower part of its allowed range.  But mass loss cannot explain the $+$22.4$\pm$1.0, and $+$35.4$\pm$7.1 ppm values for the 2010.1 and 2016.78 eruptions.  There must be some other mechanism to explain the `large' values of $\Delta P_{nova}$.  This failure of the mass-loss mechanism to explain the observed $\Delta P_{nova}$ for U Sco (and for other novae, see Schaefer 2020b) has denied my original motivation to use measured $\Delta P_{nova}$ to determine the much-needed and badly-known $M_{ejecta}$.

{\bf (2)~~}The second well-known mechanism for sudden period changes across nova events is called `frictional angular momentum loss' (FAML).  In this mechanism, during the eruption, the companion star is continuously plowing through the gas cloud formed by the expanding nova shell, and there will be dynamical friction that slows its forward velocity, with this being translated into a change in $P$.  This will move the binary into a smaller orbit, with a negative $\Delta P_{FAML}$.  This is always working against the period change from mass loss, $\Delta P_{ml}$.  With the standard calculation (e.g., Livio, Govarie, \& Ritter 1991), I get $\Delta P_{FAML}$/$P$ of -0.0072 ppm.  This value is very small because the ejecta mass for RNe is very small, say, when compared to classical novae.  FAML is always completely negligible for U Sco.

{\bf (3)~~}A third mechanism to change $P$ across a nova event is essentially the magnetic braking of the companion star's rotation due to its magnetic field trying to force co-rotation of the gas in the expanding nova shell (Martin, Livio, \& Schaefer 2011).  Detailed calculations show that the $P$ will decrease for systems with the companion star less massive than the white dwarf.  This effect is negligible in size, even compared to FAML, unless the companion star's magnetic field is very high, with a Alfv\'{e}n radius comparable to the binary's semi-major axis.  With no evidence for a highly magnetic companion star, plus the apparent accretion stream and hotspot structure in the accretion disc, this third $\Delta P_{nova}$ mechanism is negligible in size.

{\bf (4)~~}J. Frank (Louisiana State University) has pointed out a fourth $\Delta P_{nova}$ effect, where the nova eruption ejects matter asymmetrically into space, with this acting with a jet force back on the white dwarf (Schaefer et al. 2019; Schaefer 2020a).  This $\Delta P_{jet}$ can be large and negative if the nova ejecta is jetted in the forward direction of the white dwarf's orbital motion, or the $\Delta P_{jet}$ can be large and positive if the ejecta is highly asymmetric towards the backward direction.  The `jet' effect is calculated as
\begin{equation}
\frac{\Delta P_{jet}}{P} = \frac{M_{ejecta}}{(M_{WD}+M_{comp})}\left(- q\xi\frac{3 V_{\rm ejecta}}{ 2V_{WD}}\right)\, .
\end{equation}
The expansion velocity of the nova shell, $V_{ejecta}$, is seen to be 4800 km/s from the half-width zero-intensity of the H$\alpha$ line early in the eruption.  The orbital velocity of the white dwarf, $V_{WD}$, is calculated to be 109 km/s for U Sco.  The mass ratio, $q$=$M_{comp}$/$M_{WD}$, is 0.73$\pm$0.15.  Judging from the projected views of nova shells, the jets of ejecta appear to usually have very large opening angles, more like one-hemisphere having more mass than the opposite hemisphere, although at least the best observed nova shell (that around GK Per) does have a narrow jet component (Shara et al. 2012).  The asymmetry of the nova shell is represented by the parameter $\xi$.  For a shell ejection consisting of two unequal hemispheres, the value of $\xi$ can vary from $-$1 to $+$1.  (If a narrow jet is involved, then the range is from $-$2 to $+$2, depending on the orientation of the jet.)  For U Sco, $\Delta P_{jet}$/$P$ equals $-$51$\xi$ ppm.  If the U Sco white dwarf ejects all the shell as a hemisphere in the forward direction, then the fractional period change will be -51 ppm.  If the hemisphere is ejected in the backward direction, the orbital period will increase by $+$51 ppm.  Schaefer et al. (2019) reports on a small survey of resolved nova shells and finds that most have asymmetries that translate into |$\xi$| values of 0.4--0.9.  Therefore, the asymmetric ejection of the nova shell is an easy mechanism to get the observed $\Delta P_{nova}$/$P$ values.  

The first three $\Delta P_{nova}$ mechanisms are all greatly too small to explain the observed period changes, with their being no realistic chance of these being significant.  But the fourth mechanism works nicely.  This is not a proof that the U Sco period change is caused by asymmetric nova shell ejection.  But it is now the one remaining idea that has any possibility of working.

A strong advantage of the `asymmetric-shell' explanation is providing a ready understanding of why $\Delta P_{nova}$ changes greatly from eruption-to-eruption, as the `jet' randomly changes its pointing direction.  Another strong advantage of the `jet' explanation is that most of the visible nova shells show strong asymmetries, while kicks on the white dwarf are inevitable.  A third strong advantage is that this mechanism gives a simple understanding for why my 6 CN eruptions and 6 RN eruptions have a well-distributed mix with 4 eruptions with a large-negative $\Delta P_{nova}$, 3 eruptions with a near-zero $\Delta P_{nova}$, and 5 eruptions with a large-positive $\Delta P_{nova}$, which is appropriate for random pointing directions for the `jet'.  Despite these three strong evidences for the jetting, there are still open questions.  The foremost open question is the need for a realistic physical model that produces the asymmetric ejecta.  The lack of any such model in place is {\it not} an argument against the the fourth mechanism, as the well-resolved asymmetric shells expanding around many CNe already proves that a solution exists.  Still, the `jet' hypothesis does not have any direct evidence in its favor.  This problem is {\it not} an argument against the hypothesis, because I cannot think of any means to {\it test} its predictions with data in hand.

In all, the asymmetric-ejecta answer satisfies the observed $\Delta P_{nova}$ values, explains the eruption-to-eruption variability, is already known to be inevitable due to known strong asymmetries in nova shells, and satisfies the demographics of the $\Delta P_{nova}$ distribution for 12 eruptions.  Further, all other ideas to date fail  to come close.  Nevertheless, I judge that there is no proof for the asymmetric ejecta hypothesis.  Despite this, the case has much strong evidence and is now the one remaining possibility, and I further judge that the case is convincing.

\subsection{Magnetic braking is driving the steady period changes}

How can the steady period changes between eruption be explained?  To be specific, I report the $\dot{P}$ values for the four inter-eruption intervals to be $-$3.2$\pm$1.9, $-$1.1$\pm$1.1, $-$21.1$\pm$3.2, and $-$8.8$\pm$2.9, all in dimensionless units of $10^{-9}$.

The steady period changes between eruptions can only arise from the steady mass transfer due to accretion and from angular momentum losses to the binary as a whole due to effects such as magnetic braking and stellar winds.  The general equation is 
\begin{equation}
\dot{P} = [3 P \dot{M}/M_{comp})(1-q)] + [3 P \dot{J}/J].  
\end{equation}
This $\dot{P}$ value is dimensionless.  The first term is for the effects of mass transfer, and must always be positive for the mass ratio of U Sco (i.e., $P$ increases).  The second term collects the effects of a stellar wind from the companion star and the magnetic braking in that wind, with this necessarily making for a decrease of the binary's angular momentum ($J$), $\dot{J}$$\leq$0, and a $P$ decrease.

The accretion rate, $\dot{M}$, must be very high to support the $\sim$10 year recurrence time, and Shara et al. (2018) derive a value of 5$\times$10$^{-8}$ M$_{\odot}$/year.  The companion mass, $M_{comp}$, is something like 1.0$\pm$0.2, while the mass ratio, $q$ is 0.73$\pm$0.15.  The mass transfer then requires $\dot{P}$ to be close to $+$1.4$\times$10$^{-10}$ for this effect alone.  This is the base mechanism that must be operating in U Sco.  With magnetic braking, the total steady period change can only be smaller, possibly going negative.  The comparison of this required mechanism ($+$0.14 in units of $10^{-9}$) versus the observed values for the four inter-eruption intervals ($-$3.2, $-$1.1, $-$21.1, and $-$8.8) shows that the mass transfer effect is relatively small, being completely dominated by magnetic braking effects. The U Sco $\dot{P}$ is being driven by magnetic braking.

Magnetic braking has long been realized to be the dominant driver of CV evolution (Patterson 1984; Rappaport, Joss, \& Webbink 1982; Knigge, Baraffe, \& Patterson 2011).  Unfortunately, the size and scaling properties of the braking effect are only badly known.  To illustrate this, fig. 2 of Knigge et al. (2011) overplots the many published models for $\dot{J}$, with these spanning 5 order-of-magnitude, while the power-law index for the dependency on the donor mass ($\gamma$) goes from 0 to 4.  Further, unfortunately, all the models in the CV literature are applicable out to periods of only around 7 hours.  That is, none of the models used by the CV community can be used for U Sco, nor for any nova with an evolved companion star.

Unfortunately, in particular for U Sco, the estimation of the magnetic braking effect has three big problems.  First, the strength of the stellar wind coming from the companion star is completely unknown.  Second, the strength of the companion's magnetic field is only known to be not-small.  As a G-type star spun up to a rotation period of 1.23 days, a high level of magnetic activity is expected.  Further, the nearly identical RN system V2487 Oph has incredible Superflares ($\sim$10$^{39}$ erg flares once a day) that require large magnetic fields with a complex structure (Schaefer, Pagnotta, \& Zoppelt 2022).  Third, the structure of the companion's magnetic field is completely unknown, in particular, the strength of the dipole component.  The dipole strength dominates the far field that determines the size of the Alfv\'{e}n radius, within which the stellar wind is forced into co-rotation with the star, that determines the lever arm for transferring angular momentum to the wind.  With these three very big problems, $\dot{J}$ cannot be estimated from the known properties of U Sco specifically, even to within a few orders-of-magnitude.

Fortunately, magnetic braking for cases similar to U Sco has been extensively considered by the binary star community, specifically for the common RS CVn systems (e.g., Nanouris et al. 2011).  Regular RS CVn stars are binaries with the larger component being an F or G class subgiant (or main sequence star) orbiting a lower-mass main sequence star with 1$<$$P$$<$14 days.  Many RS CVn stars have the same critical properties (G0 subgiant spun up to a 1.23 day rotational period) as the companion in U Sco.  For estimating the magnetic braking in U Sco, similar RS CVn stars might provide answers.

Unfortunately, the stellar wind and magnetic braking effects for the similar RS CVn stars are badly known.  This is illustrated in Table 2 of Nanouris et al. (2011), where five different published and reputable models have the range of the braking strength varying by ten orders-of-magnitude, with the power-law index ($a$) for the scaling of $\dot{J}$ with $P$ going from 1.0 to 3.0.  This means that a general model cannot provide any useable $\dot{J}$.

I can think of only one possible means to usefully estimate the magnetic braking of U Sco, and that is to find an RS CVn system with a similar star and a well-measured $\dot{P}$.  The RS CVn system must have a star that has a spectral type near G0, a radius near 2.2 R$_{\odot}$, and a rotation period near 1.23 days.  The RS CVn star must be eclipsing (to avoid problems with changing and moving starspots) with a very long record of eclipse times (to be sure that the $O-C$ parabola is due to its evolution).  Fortunately, AR Lac nicely fits the requirements.  AR Lac is an RS CVn binary, consisting of a G2 {\rm IV} star and a K0 {\rm IV} star with a 1.98 day orbital period.  AR Lac has a wonderful $O-C$ curve with 376 eclipse times from 1900 October to present, with this displaying a distinct and highly significant parabolic component (Lu, Xiang, \& Shi 2012).  Siviero, Dallaporta, \& Munari (2006) report the hotter component to have a surface temperature of 5826 K, a mass of 1.17 M$_{\odot}$, and a radius of 1.51 R$_{\odot}$, with this being a reasonably close match to the U Sco companion star.

Lu et al. (2012) fit their 111 year $O-C$ curve to find that the parabolic term has a coefficient of ($-$2.11 $\pm$ 0.059) $\times$ 10$^{-9}$.  Equating this with $0.5P\dot{P}$ (see equation 7), $\dot{P}$ equals $-$2.128$\pm$0.060 in the dimensionless units of 10$^{-9}$.  The formal error bar is rather small, but for comparison to U Sco, the difference could well be quite large.  That is, AR Lac has no Roche lobe overflow, there are small differences in the stellar properties that I have no way to correct, the stellar winds of the two stars might well be substantially different, and the magnetic field strengths and structures might well be substantially different.  The real error bar for applying the $\dot{P}$=$-$2.1 $\times$ 10$^{-9}$ to U Sco could easily be large.  Looking around at measured period changes for other RS CVn stars, and seeing the range of typical models, I judge that the real uncertainty might be one-order-of-magnitude (i.e., a factor of 10).  This poor accuracy is still good enough for getting a reasonable idea of what is going on in the U Sco system.

For AR Lac, the $3\dot{J}/J$ equals $-$1.1 (in units of 10$^{-9}$), while this is also our best estimate for U Sco.  This is to be compared to the four measures for U Sco between eruptions ($-$3.2, $-$1.1, $-$21.1, and $-$8.8, in the same units).  Recalling that the mass transfer term is negligibly small and that the real uncertainty on the predicted $-$1.1 is large, I see that the observed values are consistent with the the magnetic braking expectations.  With no other possibility in sight, I conclude that the parabolic segments in the $O-C$ curve have the dominant cause of magnetic braking by the companion star.

An open question is the cause of the large variability in $\dot{P}$ from eruption-to-eruption.  The variation is over one order-of-magnitude from before the 2010.1 eruption to after.  With no reliable physical model, no detailed or confident answer can be known.  Still, the answer must somehow involve some combination of changing magnetic field strength and stellar wind strength.  Even so, the critical change on the companion star must have been somehow provoked by the 2010.1 nova event, as there was a sudden change in $\dot{P}$ at the time of the eruption.    And this same critical change did {\it not} occur at the time of the 1999.2 eruption.  For this, it is important that all U Sco eruptions have identical light curves and spectral development, see Schaefer (2010), and it is mysterious as to what property of the system was changed by the eruption to cause many subsequent years to have a new unchanging $\dot{P}$.  In all, it is a mystery as to why the U Sco $\dot{P}$ changes greatly, but only at the times of eruption.

\subsection{$\Delta$P and Hibernation}

The original `Hibernation' model (Shara et al. 1986) is a seductive scenario for the evolution of CVs.  Part of its original motivation was the first measure of $\Delta P_{nova}$ for a nova, BT Mon, where its $P$ was observed to {\it increase} substantially across its 1939 eruption (Schaefer \& Patterson 1981).  This Original-Hibernation model makes a variety of specific predictions that should apply to U Sco.  (Original-Hibernation uses many novae with periods above the Period Gap as supporting evidence, including the critical BT Mon, while CVs with relatively long periods and evolved companion stars, such as Nova Sco 1437 with $P$=0.53 days and HR Lyr with $P$=0.905 days,  are included as highlighted `proof' of hibernation.)  Here I will discuss just two of the predictions of the original Hibernation model that have failed completely for U Sco:  {\bf (1)~}  The hallmark prediction of Hibernation is that the brightness level of the post-nova will fade fast and far as the newly-detached system turns off its accretion flow.  U Sco is nearly the fastest known nova, with a duration of 60 days in the optical and X-ray, and any irradiation should be stopped long before the next eruption, therefore this predicted effect should be apparent.  To the contrary, U Sco does {\it not} fade after the eruption and irradiation are stopped (see Fig. 2).  Indeed, after the end of the 2010.1 eruption, the system actually was substantially {\it brighter} than before the eruption.  Thus, the hallmark prediction of original Hibernation has failed for U Sco.  {\bf (2)~}The entire reason and mechanism driving Hibernation is that $P$ has a sudden and large increase, driving the binary orbit apart, and Hibernation requires that $\Delta P_{nova}$ be very large and positive.  Schaefer (2020a) calculates the required period change to produce even mild hibernation for U Sco as $\Delta P_{nova}$/$P$~$>$~$+$1380 parts-per-million (ppm).  To the contrary, U Sco has $\Delta P_{nova}$/$P$ values of $+$3.9$\pm$6.2, $+$22.4$\pm$1.0, and $+$35.4$\pm$7.1 ppm for the three measured eruptions.  That is, even the largest period change in U Sco is $>$40$\times$ smaller than required by Hibernation.  In all, U Sco is a counterexample for Original-Hibernation.

Starting with Hillman et al. (2020), the Original-Hibernation model has changed drastically to what I am labeling as the `New-Hibernation' model.  The new version uses all the same physics and mechanism, but the new exhaustive calculations demonstrate that CVs with periods above the Period Gap should {\it not} display any hibernation at all.  That is, for 94.6 per cent of all known nova systems (i.e., all those with $P$$>$3 hours), there will be no fading of brightness or accretion between eruptions, and there will be no apparent increase in $P$ across eruptions.  With this, the New-Hibernation model is not applicable to U Sco, or a better statement is that the hibernation scenario predicts exactly the same outcome as any other generic model where nothing happens.  The case of U Sco provides a strong counterexample for Original-Hibernation, but is not covered by New-Hibernation.

The New-Hibernation scenario might not apply to U Sco, but the U Sco measures do apply for understanding the underlying physical mechanism for $\Delta P_{nova}$ that is the driver of New-Hibernation.  Here are two problems for the hibernation mechanism in general as demonstrated with U Sco:  {\bf (1)~}The U Sco measures of $\Delta P_{nova}$/$P$ are near-zero and very small when compared to that required to make even mild hibernation, and this is suggestive that the real combination of period-change mechanisms in CVs is unlikely to be large enough.  This dilemma is not just for U Sco, but is also seen in 6 ordinary classical novae (BT Mon, QZ Aur, RR Pic, DQ Her, HR Del, and V1017 Sgr) for which I have measured $\Delta P_{nova}$ (Schaefer 2020b).  The most-positive $\Delta P_{nova}$ is for BT Mon that is still 40$\times$ too small to make for mild hibernation, while five-out-of-six have {\it negative} $\Delta P_{nova}$.  The negative $\Delta P_{nova}$ is impossible for any version of Hibernation.  Indeed, these are empirical demonstrations for what could be called `anti-Hibernation'.  Further, T Pyx is below the Period Gap, New-Hibernation should be applicable, yet my measured $\Delta P_{nova}$/$P$ = $+$48.9$\pm$7.9 ppm is greatly too small to allow for any hibernation.  {\bf (2)~}U Sco has its $\Delta P_{nova}$ change greatly and significantly from eruption-to-eruption, and such is not possible from any of the published mechanisms for period changes during a nova eruption.  That is, the standard mechanisms for $\Delta P_{nova}$, the driving force in New-Hibernation, are greatly smaller than the observed effect, and there must be some dominant physical mechanism that is missing from New-Hibernation (see Section 5.2).  It does not matter whether the missing mechanism is asymmetric-ejecta or not, rather the point is that New-Hibernation does not include a physical effect that dwarfs its hallmark mechanism of $\Delta P_{ml}$.  With this, New-Hibernation cannot have results with any degree of reliability.


\subsection{What is the evolution of U Sco?}

The evolution of CVs is now the forefront issue for CV research.  I am aware of little literature that is relevant for understanding the evolution of U Sco and T CrB.  The general case of the past and future for CVs with subgiant and red giant companions has been largely ignored.  And this is a lot larger problem than simply understanding the famous RNe U Sco, T CrB, and RS Oph.  After an exhaustive analysis of new photometric and spectral energy distribution data, I have a comprehensive list of 156 reliable nova periods, for which 31 per cent have an evolved companion.  This fraction is greatly larger than most workers realize.  Here, I will do my best to understand the past, present, and future evolution of U Sco. 

\subsubsection{What is the evolution up to the present date?}

Even though not in any literature, the basic outline for the formation of U Sco (and all CVs with evolved companions) is clear.  It all starts with a wide binary with stellar masses something like 2--8 M$_{\odot}$ and perhaps 1--3 M$_{\odot}$.  The more massive star evolves off the main sequence, grows large, has interactions with the secondary, and settles down as a white dwarf.  This is a common path of binary evolution, and ordinary evolution leaves a $\sim$1 M$_{\odot}$ secondary in orbit around a white dwarf.  The fate of this binary depends to a strong degree on the orbital period at the time of the white dwarf formation.  If $P$ is less than something like 10 hours, the orbit will eventually get ground down by magnetic braking, the secondary star will come into contact, gas will flow through the inner Lagrange point, an accretion disc will form, and an ordinary CV will be seen.  Alternatively, for $P$$>$10 hours, or some such limit, the orbit cannot in-spiral on any useful time-scale.  For these systems, the next stage is set when the secondary star starts expanding as it evolves off the main sequence.  An isolated star of 1 M$_{\odot}$ will take nearly the age of the galaxy to start off the main sequence, whereas a secondary in such a binary can have suffered mass striping while the primary was a red giant, and its nuclear burning clock can be farther along than is apparent for the current mass of the secondary.  (There must be some means of taking mass off the secondary, as for example, the companion stars in T CrB and RS Oph are $\sim$0.7 M$_{\odot}$, and such stars could not have had time to evolve to become a large red giant without having its nuclear burning `clock' running fast when the star had a substantially higher mass.)  When the secondary expands as it leaves the main sequence, it will fill its Roche lobe, initiate mass transfer, form an accretion disc, and appear as a CV with an evolved companion.

If the original $P$ for the binary just after the white dwarf formed was from roughly 10 hours to 10 days, then the secondary star will come into contact as a subgiant star.  This would be the case for U Sco, where its progenitor would have been a white dwarf plus main sequence star with a period of around 1.23 days.  Once this progenitor comes into contact, the binary is like U Sco.  If the original $P$ was roughly from 10 to 2000 days, then the secondary star would have to expand all the way to a red giant stage before contact.  This would be the case for T CrB and RS Oph, the only well-known novae with red giant companions.

\subsubsection{Is the white dwarf gaining or losing mass?}

Are the white dwarfs in CVs gaining or losing mass?  This is a tricky and contentious problem.  At stake is whether novae and RNe have $M_{WD}$ increasing to the Chandrasekhar limit, ultimately resulting in a Type Ia supernova.  With novae possibly changing into RNe and then Type Ia supernovae, this channel has long been one of the favorite single-degenerate models in the  top-level important Supernova Progenitor problem.  The case of U Sco is often pointed out as a progenitor, because after all, it certainly has a white dwarf already very close to the Chandrasekhar mass and material is piling on to it at a very high rate, so, simplistically, it is only a matter of $\sim$10$^5$ years before it explodes.

The issue of gaining/losing mass is a balance between the mass ejected ($M_{ejecta}$) over an eruption cycle of length $\delta Y$ versus the mass accreted over the eruption cycle ($\delta Y$$\langle \dot{M} \rangle$).  For the mass accreted, we have three reasonable inputs.  First, a reasonable accretion rate from the absolute magnitude ($M_q$=1.58$\pm$0.52) is $\dot{M}$=9$\times$10$^{-8}$ M$_{\odot}$/year, which for $\delta Y$=10 years has the accumulated mass of 9$\times$10$^{-7}$ M$_{\odot}$.  Second, based on the eruption amplitude and decline time, Shara et al. (2018) calculated an accretion rate of 5.1$\times$10$^{-8}$ M$_{\odot}$/year, and a trigger threshold of 5.3$\times$10$^{-7}$ M$_{\odot}$.  Third, the theoretical calculation of the trigger conditions have M$_{trigger}$ is near 16$\times$10$^{-7}$ M$_{\odot}$ (Shen \& Bildsten 2009).  From the scatter in these estimates, $\delta Y$$\langle \dot{M} \rangle$ is known to within a factor of three or so.

The big question then comes down to whether M$_{ejecta}$ is greater than or less than $\sim$(5--16)$\times$10$^{-7}$ M$_{\odot}$.  Unfortunately, for this, the real uncertainties, both observational and theoretical, are many orders of magnitude.  Appendix A of Schaefer (2011) documents that the traditional methods for measuring M$_{ejecta}$ has six independent causes for large errors, with uncertainties of 1500$\times$, 100$\times$, 4$\times$, $\sim$10$\times$, 1000$\times$, and $\gtrsim$8$\times$ respectively, all usually ignored.  Values published for U Sco specifically vary over the range (0.01--7.2) $\times$ 10$^{-7}$ M$_{\odot}$.  And my hopeful method to measure M$_{ejecta}$ from the $\Delta P_{nova}$ has failed completely because the sudden period change is dominated by a poorly known mechanism that is one-to-two orders-of-magnitude larger (see Section 5.2).  Theory provides no useful guidance, as four models specific to U Sco predict M$_{ejecta}$ scattered evenly over a range of (2.1--44) $\times$ 10$^{-7}$ M$_{\odot}$ (Appendix A of Schaefer 2011).  With such huge uncertainties in both observations and theory, there can be no confidence in deciding whether M$_{ejecta}$ is greater than or less than $\sim$(5--16)$\times$10$^{-7}$ M$_{\odot}$.

The question of whether U Sco is gaining or losing mass can also be approached by considering the elemental composition of the nova ejecta.  {\it If} U Sco is a neon nova (i.e., with the ejecta containing a significant excess of the element neon, beyond that possible from the nuclear burning), then the only way to get the neon is by dredging up the surface of an ONeMg white dwarf.  With the ejection of dredged up material during each eruption, the white dwarf must be losing mass each eruption.  Such a system will evolve with a long whittling away of the mass of the two stellar components, with lengthening nova recurrence timescale and shortening orbital period, ultimately to quietly fade away.  Unfortunately, the case for U Sco as a neon nova is confused.  Mason (2011) claimed that U Sco is a neon nova, based on the strong neon lines (compared to the [O{\rm III}] $\lambda$5007 line) and on the P-Cygni profile for ultraviolet lines as characteristic only of neon novae.  But then in a corrigendum, Mason (2013) pointed out a case (with a high electron density of $>$10$^6$ cm$^{-3}$) where the neon abundance need not be more than mildly enriched.  This point was immediately denied by their collection of measures for the best cases of neon novae, which have strong neon lines (compared to the oxygen lines) yet have electron densities $>$10$^6$ cm$^{-3}$.  Further, I do not see how the U Sco ejecta can have such high electron densities as $>$10$^6$ cm$^{-3}$ at 104 days past peak (of the 2010 eruption) with its very low ejecta mass and very high expansion velocity.  So we are left with U Sco having very bright neon emission lines and P-Cygni profiles in the ultraviolet, looking just like the known neon novae.  Still, I cannot see that a confident conclusion has been reached.  In all, based on the composition of the ejecta, we are left with no-conclusion as to whether the U Sco white dwarf is gaining or losing mass.

\subsubsection{What is driving the accretion?}

U Sco has an incredibly high accretion rate of $\sim$(5--16)$\times$10$^{-7}$ M$_{\odot}$/year, which places it near the maximum possible for any CV, to avoid the steady-hydrogen-burning regime.  What is driving this rare and extreme accretion?  The high accretion rate can only be caused by a rapid shrinkage of the companion's Roche lobe size or by a rapid expansion of the companion star itself.

The ordinary expansion of a star after leaving the main sequence is well known from tracks of models in the H-R diagram.  With no accurate current mass for the companion, and no useful estimate of its original main-sequence mass, the expansion rates cannot be determined with any high accuracy.  For a characteristic path in the middle of the subgiant track, a 1.0 M$_{\odot}$ star will expand from 1.75 to 2.47 R$_{\odot}$ in 0.5 Gyr, for an expansion velocity of 1.0 m/year.  Such a slow expansion cannot provide the observed accretion rate for any plausible gas density along the surface of the Roche lobe, by many orders-of-magnitude.  Choosing any other plausible evolutionary track still has the companion's expansion rate too small to matter by many orders-of-magnitude.  This demonstrates that the expansion of the companion star is not contributing to driving the accretion.

The only other possibility is that the companion's Roche lobe is shrinking at some fast rate.  A changing size of the Roche lobe is tied to a proportionate change of the orbital period ($\dot{R}_{Roche}/R_{Roche}$=$\frac{2}{3}$$\dot{P} / P$).  For U Sco, an apparently good measure of the evolutionary $\dot{P}$ is ($+$0.20$\pm$0.05)$\times$10$^{-9}$ from 1945--2022, averaging over seven eruptions.  But the trouble is that this long-term evolutionary trend is {\it positive}, with the period {\it increasing}, and the companion's Roche lobe getting {\it larger}.  With such evolution, U Sco cannot have any sort of a high $\dot{M}$.  

Something is driving the high $\dot{M}$ in U Sco, and it can only be the rapid shrinkage of the companion's Roche lobe.  Thus, the evolutionary $\dot{P}$ must be negative, with an overall smoothed parabola that is concave-down in Fig. 9.  Something is wrong with the upturn in the uniform evolution prominently seen in Fig. 9.  Perhaps chaotic variations in $\Delta P$ and $\dot{P}$ only average out to some evolutionary average rate on time-scales $\gg$77 years and $\gg$7 eruptions.

Presumably, the long-term average $\dot{P}$ will not be much greater than the maximal period change observed during the inter-eruption intervals.  That is, the evolutionary $\dot{P}$ cannot be expected to be more negative than the $\dot{P}$ from the 2010.1--2016.78 interval, $-$2.1$\times$10$^{-8}$.  For this maximal evolutionary rate and a 2.2 R$_{\odot}$ companion, the companion's Roche lobe would be contracting at a rate of 6.3 km/year.  This is to be compared to an atmospheric scale height of order 700 km.  With such a rate, the Roche lobe would have moved inward by close to one atmospheric scale height in the 116 years going back to 1906, the 1906 accretion would have been near a factor of 2.7$\times$ smaller than currently, and the oldest $\delta Y$ values should be more like 30 years.  This is clearly contradictory to the eruption record, and the Roche lobe is not shrinking at a rate anywhere near to 6.3 km/year.

Another perhaps severe problem arises when trying to get accretion of $\sim$1$\times$10$^{-7}$ M$_{\odot}$/year when just 6.3 km of shell crosses over the Roche lobe each year.  To get the observed rate, the density in the accretion zone must be roughly 0.001 gm/cc.  This does not sound like a dense atmosphere by terrestrial standards, but for a 1 M$_{\odot}$ star at 2.2 R$_{\odot}$, such a density only occurs far inside the expected surface.  For a similar isolated star, the 0.001 gm/cc density level occurs roughly 5 per cent inward from the photosphere.  That is, the Roche lobe has shrunk far inside the companion's surface, and the companion star has not responded by having local expansion to achieve any condition close to hydrostatic equilibrium.  The restoration of hydrostatic equilibrium is on the sound-crossing time-scale, which is much faster than the time-scale for the contraction of the Roche lobe.  I do not understand how it is possible for the stellar atmosphere to have such a high density near the Roche lobe.  The vertical density structure near the Roche lobe of a fast accreting system is a difficult problem for current astrophysics.  Still perhaps a solution can be found where U Sco can have its high $\dot{M}$ driven by some hypothetical maximal period change.  Until then, the mechanism that is driving the very high accretion in U Sco remains mysterious.

\subsubsection{Overview on evolution}

The evolution of CVs is now the forefront issue for CV research, and is also the higher-level goal of my program to measure eclipse times and period changes of U Sco and other novae.  CV evolution is driven by period changes.  The U Sco case shows that the evolution is controlled by {\it both} $\Delta P_{nova}$ and $\dot{P}$, with these roughly cancelling out.  Both $\Delta P_{nova}$ and $\dot{P}$ show large variations from eruption to eruption, with the causes being mysterious.  The physical mechanism for the sudden period changes across the nova events cannot be any of the well-known mechanisms (e.g., mass loss, frictional angular momentum loss), but instead is apparently caused by the ubiquitous asymmetric ejection of mass into the nova shell pushing a jet force on to the white dwarf.  The physical mechanism for the steady period change between eruptions is not from the steady mass transfer, but rather must be dominated by the poorly-known mechanism of magnetic braking in a stellar wind of the companion star.

The very-high accretion rate in U Sco is not being driven by the ordinary evolutionary expansion of the subgiant companion star (such is greatly too small to matter in all cases), hence the driving for the very-high $\dot{M}$ can only come from a fast shrinkage of the Roche lobe and by a large {\it negative} $\dot{P}$.  Unfortunately, the 77 year $O-C$ records shows that $\dot{P}$ is {\it positive} over 7 eruption cycles.  I can only explain this discrepancy by suggesting that the vagaries of the variations in $\Delta P_{nova}$ and $\dot{P}$ randomly happened to produce a positive $\dot{P}$ over the 1945--2022 interval, whereas intervals $\gg$77 years would produce a steady average negative $\dot{P}$.  For any negative $\dot{P}$, the gas density at the Roche surface is, apparently, impossibly high to allow the observed $\dot{M}$.  There is no understanding of the cause or mechanism for the very-high observed accretion rate in U Sco.

For U Sco (and the 31 per cent of nova systems with evolved companions), the progenitor system was a wide detached binary with a white dwarf and a $\sim$1 M$_{\odot}$ secondary star.  The CV was formed only after ordinary evolution of the secondary star had it leave the main sequence and expand until coming into contact with its Roche lobe.  U Sco now has a very high mass white dwarf (1.36 M$_{\odot}$) and a very high accretion rate ($\sim$(5--16)$\times$10$^{-7}$ M$_{\odot}$/year), with both of these conditions required to make the system a recurrent nova (with $\delta Y$$\sim$10 years).  Whether the white dwarf is gaining or losing mass is not known, with order-of-magnitude uncertainties in the balance between accretion and ejection.  In the end, the fate of U Sco is still a mystery.

\section{ACKNOWLEDGEMENTS}

I am grateful for the eclipse time series observations of G. Myers and W. Cooney, with these being critical for recognizing and understanding the 2016.78 eruption.  E. Christensen and K. Battams have kindly passed along CSS and LASCO images.  Further, many observers and data-handlers provided light curves from the AAVSO, CRTS, Pan-STARRS, DECam, {\it Kepler} satellite, ATLAS, and ZTF.

\section{DATA AVAILABILITY}

All of the photometry data are explicitly given in Tables 2, 3, 4, and 6, or are publicly available from the references and links in Table 1.


{}

\bsp	
\label{lastpage}
\end{document}